\documentclass[aps,preprint,nofootinbib,eqsecnum]{revtex4}%
\usepackage{amsfonts}
\usepackage{amsmath}
\usepackage{amssymb}
\usepackage{graphicx}%
\providecommand{\U}[1]{\protect\rule{.1in}{.1in}}

\begin{document}
\title{}
\author{}
\maketitle

\leftline {USC-08/HEP-B4 \hfill }{}{\vskip-1cm}

{\vskip 2cm}

\begin{center}
{\Large \textbf{Relativistic Harmonic Oscillator Revisited}}\footnote{This
work was partially supported by the US Department of Energy under grant number
DE-FG03-84ER40168.}{\Large \textbf{\ }}

{\vskip1.0cm}

\textbf{Itzhak Bars}

{\vskip1.0cm}

\textsl{Department of Physics and Astronomy}

\textsl{University of Southern California,\ Los Angeles, CA 90089-2535 USA}

{\vskip2cm} \textbf{Abstract}
\end{center}

The familiar Fock space commonly used to describe the relativistic harmonic
oscillator, for example as part of string theory, is insufficient to describe
all the states of the relativistic oscillator. We find that there are three
different vacua leading to three disconnected Fock sectors, all constructed
with the same creation-annihilation operators. These have different spacetime
geometric properties as well as different algebraic symmetry properties or
different quantum numbers. Two of these Fock spaces include negative norm
ghosts (as in string theory) while the third one is completely free of ghosts.
We discuss a gauge symmetry in a worldline theory approach that supplies
appropriate constraints to remove all the ghosts from all Fock sectors of the
single oscillator. The resulting ghost free quantum spectrum in d+1 dimensions
is then classified in unitary representations of the Lorentz group SO(d,1).
Moreover all states of the single oscillator put together make up a single
infinite dimensional unitary representation of a hidden global symmetry
SU(d,1), whose Casimir eigenvalues are computed. Possible applications of
these new results in string theory and other areas of physics and mathematics
are briefly mentioned.

\newpage

\tableofcontents

\newpage

\section{Introduction}

\label{intro}

The relativistic harmonic oscillator in $d$ space and $1$ time dimensions that
will be discussed in this paper is the straightforward generalization of the
non-relativistic case by replacing position and momentum by their relativistic
counterparts $x^{\mu},p^{\mu}\;$as SO$\left(  d,1\right)  \;$vectors.

There is a long history of studies of the relativistic harmonic oscillator.
Some of these were motivated by possible physical applications of the
relativistic oscillator as an \textquotedblleft imperfect
model\textquotedblright\ \cite{Feynman}\footnote{Feynman called this approach
an imperfect model. Indeed, as is now known, the physically correct
description of systems such as quark-antiquark bound states is formulated in
the context of quantum chromo-dynamics. Approximations to chromo-dynamics for
slow moving heavy quarks is handled in terms of a non-relativistic potential
$V\left(  \vec{r}\right)  =\alpha\left\vert \vec{r}\right\vert -\beta
/\left\vert \vec{r}\right\vert ,$ rather than the relativistic oscillator,
while for fast moving light quarks this approach is not an accurate model.} to
approximate bound states of quarks in a relativistic setting. This involved
solving the relativistic oscillator eigenvalue equation\footnote{We absorb all
dimensionful parameters as well as the frequency of the oscillator by
rescaling the $x^{\mu},p^{\mu}$.} in the space of the relative coordinate
$x^{\mu}=x_{1}^{\mu}-x_{2}^{\mu}$
\begin{equation}
\frac{1}{2}\left(  -\partial^{\mu}\partial_{\mu}+x^{\mu}x_{\mu}\right)
\psi_{\lambda}\left(  x\right)  =\lambda\psi_{\lambda}\left(  x\right)  ,
\label{ho}%
\end{equation}
and associating the eigenvalue $\lambda$ with the mass of the bound state.

Some solutions of this equation appeared in earlier papers \cite{yukawa}%
\cite{dirac} and in follow up applications \cite{kim1}, but the Lorentz
symmetry properties of these solutions remained obscure to this day
\cite{kim2}. Lorentz covariant solutions based on a vacuum state $\psi
_{vac}\left(  x\right)  \sim\exp\left(  -x^{\mu}x_{\mu}/2\right)  $ that is a
Lorentz invariant Gaussian have a number of problems, including issues of
infinite norm and negative norm states, that were suppressed with \textit{ad
hoc} arguments for the sake of going forward with the physical application
\cite{Feynman}. More careful analyses, that paid attention to Lorentz
properties by using infinite dimensional unitary representations of SO$(3,1)$
\cite{naimark} relevant for this problem \cite{joos}\cite{zmud}, suggest that
there are solutions of this equation in different spacelike and timelike
patches that should be matched across the lightcone $x^{\mu}x_{\mu}=0.$
Several examples of this covariant approach using generalized relativistically
invariant potentials $V\left(  x^{\mu}x_{\mu}\right)  $ that may be different
in different patches were also studied \cite{blaha}. Proposals to confine the
solutions to only \textit{part} of the spacelike region were also discussed
\cite{horwitz}\cite{land}.

It is fair to say that there remains open questions regarding the symmetry
properties of the solutions of this differential equation. Understanding the
symmetry properties of the solutions will be the focus of the present paper.

The same equation arises as a building block in string theory. The phase space
$X^{\mu}\left(  \tau,\sigma\right)  $, $P^{\mu}\left(  \tau,\sigma\right)  $
of an open relativistic string can be expressed in terms of its normal modes
\begin{equation}
X^{\mu}=x_{0}^{\mu}\left(  \tau\right)  +\sqrt{2}\sum_{n=1}^{\infty}x_{n}%
^{\mu}\left(  \tau\right)  \cos\left(  n\sigma\right)  ,\;\;P^{\mu}=\frac
{1}{\pi}p_{0}^{\mu}\left(  \tau\right)  +\frac{\sqrt{2}}{\pi}\sum
_{n=1}^{\infty}p_{n}^{\mu}\left(  \tau\right)  \cos\left(  n\sigma\right)
\end{equation}
Except for the center of mass mode $\left(  x_{0}^{\mu},p_{0}^{\mu}\right)  $
that behaves like a free particle, the normal modes $\left(  x_{n}^{\mu}%
,p_{n}^{\mu}\right)  $ are relativistic harmonic oscillator modes with
frequency $\omega_{n}=n.$ The quantum wavefunction of a string in position
space depends on all of these modes
\begin{equation}
\psi\left(  X^{\mu}\right)  =\psi\left(  x_{0}^{\mu},x_{1}^{\mu},x_{2}^{\mu
},\cdots\right)  .
\end{equation}
This is the string field that appears in string field theory \cite{witten}%
\cite{msft}. It obeys a differential equation $\left(  L_{0}-1\right)
\psi\left(  X^{\mu}\right)  =0$ where $L_{0}$ is the zeroth Virasoro operator
which is basically a sum of operators $Q_{n}=\frac{1}{2}\left(  p_{n}%
^{2}+n^{2}x_{n}^{2}\right)  $ of the type that appears in Eq.(\ref{ho}%
)\footnote{The constant $a=\frac{1}{2}\left(  d+1\right)  \sum_{n}n$ subtracts
the vacuum energy of all the oscillators. After this renormalization the
Virasoro constraint is determined as $L_{0}=1.$ \label{a}}
\begin{equation}
L_{0}=-\partial_{0}^{\mu}\partial_{0\mu}+\sum_{n=1}^{\infty}\frac{1}{2}\left(
-\partial_{n}^{\mu}\partial_{n\mu}+n^{2}x_{n}^{\mu}x_{n\mu}\right)  -a.
\end{equation}
If this had been the only equation for the string field $\psi\left(  X^{\mu
}\right)  $, then the solution would have been a direct product of solutions
of Eq.(\ref{ho}) with a restriction on the sum of the eigenvalues%
\begin{equation}
\psi\left(  X^{\mu}\right)  \sim e^{ik\cdot x_{0}}%
%TCIMACRO{\dprod \limits_{n=1}^{\infty}}%
%BeginExpansion
{\displaystyle\prod\limits_{n=1}^{\infty}}
%EndExpansion
\psi_{\lambda_{n}}\left(  x_{n}\right)  ,\;\sum_{n=1}^{\infty}\lambda
_{n}=\left(  1-k^{2}\right)  . \label{fieldsln}%
\end{equation}
Here the center of mass momentum $k^{\mu}$ gives the mass-squared of the
relativistic string state $M^{2}\equiv-k^{2}=k_{0}^{2}-\vec{k}^{2}$. However,
$\psi\left(  X^{\mu}\right)  $ must also obey the Virasoro constraints
$L_{n}\psi\left(  X^{\mu}\right)  =0$. Therefore solutions for the free string
field $\psi\left(  X^{\mu}\right)  $ are linear combinations of
(\ref{fieldsln}) with different $\lambda_{n}$'s that satisfy the same mass
level, taken with coefficients such that the Virasoro constraints are also
obeyed. Such solutions were obtained in the covariant quantization approach,
which also provided a proof of the absence of negative norm ghosts in string
theory \cite{goddard}-\cite{thorn}.

As will be explained in section (\ref{symmvac}), upon a closer examination it
becomes evident that the relativistic Fock space treatment of string theory
\cite{gsw} inadvertently specializes to only the \textit{spacelike} sector of
every normal mode without any warning, namely
\begin{equation}
x_{n}^{\mu}x_{n\mu}\geq0\text{ and }p_{n}^{\mu}p_{n\mu}\geq0\;\text{for every
string mode }n\geq1\text{.} \label{spacelike}%
\end{equation}
This can give only non-negative eigenvalues $\lambda_{n}\geq0,$ and hence
Eq.(\ref{fieldsln}) is solved for $k^{2}$ by mostly timelike center of mass
momenta $k^{\mu}k_{\mu}<0,$ or positive $M^{2}.$ The exception is the tachyon
state that is forced to have spacelike momentum $k^{\mu}$ when all
$\lambda_{n}=0,$ and hence $M^{2}=-k^{2}=-1$ gives a tachyon%
\begin{equation}
\psi\left(  X^{\mu}\right)  \sim\langle X|0,k\rangle\sim e^{ik\cdot x_{0}}%
\exp\left(  -\frac{1}{2}\sum_{n=1}^{\infty}nx_{n}^{\mu}x_{n\mu}\right)
\label{tachyon}%
\end{equation}
when all string modes $x_{n}^{\mu}$ are in the spacelike region. For excited
levels this expression is multiplied by polynomials in the various $x_{n}%
^{\mu}.$

In view of the fact that the single oscillator equation (\ref{ho}) has
solutions in different spacetime regions as indicated above, a natural
question arises of whether there might be more general solutions to string
theory beyond the spacelike region of Eq.(\ref{spacelike}). This is not an
easy question to answer, both because there are the Virasoro constraints to
deal with, and because there is still obscurity in the previously known
solutions of the relativistic oscillator equations (\ref{ho}).

This bring us to the main topic of the current paper. We will investigate the
single relativistic oscillator without prejudice as to its possible physical
applications. Our main interest is to clarify the symmetry and unitarity or
lack thereof of its various solutions in various parts of spacetime. At the
end we will point out possible applications of our findings.

Our key observations will follow from hidden symmetries not discussed before.
First we point out that the symmetries of Eq.(\ref{ho}) go beyond the Lorentz
symmetry SO$\left(  d,1\right)  .$ There is a hidden symmetry SU$\left(
d,1\right)  $ that includes SO$\left(  d,1\right)  ,$ and therefore all
solutions, unitary or non-unitary, must fall into irreducible representations
of SU$\left(  d,1\right)  .$ Apparently this was never explored in previous
investigations of Eq.(\ref{ho}).

After clarifying the symmetry aspects we will build three different Fock
spaces by using the same relativistic harmonic oscillator
creation-annihilation operators. This includes a spacelike, timelike and mixed
spacetime sectors that are distinct from each other. While the spacelike or
timelike sectors have negative norm states, the mixed case is completely free
of negative norm ghosts and is covariant under SO$\left(  d,1\right)  $ and
SU$\left(  d,1\right)  $ in infinite dimensional unitary representations.
There may be more solutions in more intricate spacetime sectors than those
described in this paper, but we will not attempt to investigate them here (see
comments following Eq.(\ref{M}) and footnote (\ref{regions})).

For the single harmonic oscillator we will also discuss a worldline gauge
symmetry that removes ghosts and thereby introduces a constraint. The
covariant quantization of this constrained model is in agreement with the
general discussion. On the other hand, a gauge fixed quantization does not
capture all the sectors but is in agreement with the sectors describable in
that gauge. This simple example illustrates how a gauge fixed theory can fail
to capture all the gauge invariant sectors of a gauge invariant
theory\footnote{Another example is that the usual treatment of the lightcone
gauge in string theory fails to capture the folded string sectors of string
theory \cite{bbhp1}-\cite{folds2}. \label{foldss}}.

The new phenomena uncovered here both in the covariant quantization as well as
the gauge fixed quantization of the relativistic oscillator may provide tools
and rekindled interest to revisit string theory.

\section{Relativistic harmonic oscillator and SU$\left(  d,1\right)  $}

\label{oscillatorApproach}

For the sake of clarity, parts of our presentation, including this section,
will include some material that may be quite familiar to many readers, but
this will be compensated by simple observations that are not that familiar.

The operator $Q=\frac{1}{2}\left(  p\cdot p+x\cdot x\right)  $ which is being
diagonalized, $Q\psi_{\lambda}=\lambda\psi_{\lambda},$ can be written as usual
in terms of Lorentz covariant oscillators%
\begin{equation}
a_{\mu}=\frac{1}{\sqrt{2}}\left(  x_{\mu}+ip_{\mu}\right)  ,\;\;\bar{a}_{\mu
}=\frac{1}{\sqrt{2}}\left(  x_{\mu}-ip_{\mu}\right)  . \label{abara}%
\end{equation}
The covariant quantization rules
\begin{equation}
\left[  x_{\mu},p_{\nu}\right]  =i\eta_{\mu\nu},
\end{equation}
with the SO$\left(  d,1\right)  $ Minkowski metric $\eta_{\mu\nu},$ lead to
the relativistic quantum oscillator commutation rules
\begin{equation}
\left[  a_{\mu},\bar{a}_{\nu}\right]  =\eta_{\mu\nu}=diag\left(
-1,1,1,\cdots,1\right)  .
\end{equation}

In a unitary Hilbert space the operators $x_{\mu},p_{\mu}$ are Hermitian; in
that case $\bar{a}_{\mu}$ is the Hermitian conjugate of $a_{\mu},$ i.e.
$\bar{a}_{\mu}=\left(  a_{\mu}\right)  ^{\dagger}.$ A unitary Hilbert space
without ghosts (negative norm states) is possible only and only if $x_{\mu
},p_{\mu}$ are hermitian or equivalently if $\bar{a}_{\mu}=\left(  a_{\mu
}\right)  ^{\dagger}.$

In what follows we will seek unitary Hilbert spaces, but along the way we also
come across non-unitary Fock spaces in which $\bar{a}_{\mu}\neq\left(  a_{\mu
}\right)  ^{\dagger}.$ Therefore we prefer the more general notation $\bar
{a}_{\mu}$ in order not to confuse it with the hermitian conjugate of $a_{\mu
}$ when such vector spaces arise.

In terms of $a_{\mu},\bar{a}_{\mu}$ the operator $Q$ takes the form
\begin{equation}
Q=\frac{1}{2}\left(  p\cdot p+x\cdot x\right)  =\bar{a}\cdot a+\frac{d+1}%
{2}=a\cdot\bar{a}-\frac{d+1}{2}.\label{H}%
\end{equation}
This operator $Q$ has a larger symmetry than the evident Lorentz symmetry of
the dot products $\bar{a}\cdot a=\eta^{\mu\nu}\bar{a}_{\mu}a_{\nu}$. The
hidden symmetry is U$\left(  d,1\right)  $ whose generators are%
\begin{equation}
U\left(  d,1\right)  \text{ generators:\ }\bar{a}_{\mu}a_{\nu}.
\end{equation}
All of these $(d+1$)$^{2}$ generators commute with $Q$%
\begin{equation}
\left[  Q,\bar{a}_{\mu}a_{\nu}\right]  =\left[  \bar{a}\cdot a,\bar{a}_{\mu
}a_{\nu}\right]  =0,\label{Ud1}%
\end{equation}
hence $Q$ has U$\left(  d,1\right)  $ symmetry, and the spectrum of $Q,$
whether unitary or non-unitary, must be classified as irreducible
representations of U$\left(  d,1\right)  =$SU$\left(  d,1\right)  \times
$U$\left(  1\right)  $ unless the symmetry is broken by boundary
conditions\footnote{See the last paragraph of the Appendix for an example of
how the SU$\left(  d,1\right)  $ symmetry is broken to SO$\left(  d,1\right)
$ in the purely spacelike sector. }. The U$\left(  1\right)  $ part is just
the number operator $J_{0}$
\begin{equation}
J_{0}\equiv\bar{a}\cdot a=a\cdot\bar{a}-\left(  d+1\right)  ,\label{J0}%
\end{equation}
which is essentially the operator $Q$ up to a shift. Therefore the non-trivial
part is SU$\left(  d,1\right)  $ with $(d+1$)$^{2}-1$ generators that
correspond to the traceless tensor%
\begin{equation}
J_{\mu\nu}=\left(  \bar{a}_{\mu}a_{\nu}-\frac{1}{d+1}\eta_{\mu\nu}\bar{a}\cdot
a\right)  =\left(  a_{\nu}\bar{a}_{\mu}-\frac{1}{d+1}\eta_{\mu\nu}a\cdot
\bar{a}\right)  \label{J}%
\end{equation}
that satisfies $\eta^{\mu\nu}J_{\mu\nu}=0.$ The Lorentz generators $L_{\mu\nu
}$ for SO$\left(  d,1\right)  $ correspond to the antisymmetric part of the
tensor $J_{\mu\nu}$
\begin{equation}
L_{\mu\nu}=x_{\mu}p_{\nu}-x_{\nu}p_{\mu}=-i\left(  \bar{a}_{\mu}a_{\nu}%
-\bar{a}_{\nu}a_{\mu}\right)  =-i\left(  a_{\nu}\bar{a}_{\mu}-a_{\mu}\bar
{a}_{\nu}\right)  .\label{L}%
\end{equation}
The $L_{\mu\nu}$ are hermitian by construction as long as $x_{\mu},p_{\mu}$
are hermitian. So a unitary representation of the Lorentz group will be
obtained if and only if $\bar{a}_{\mu}=\left(  a_{\mu}\right)  ^{\dagger}.$ We
know that unitary representations of non-compact groups are infinite
dimensional except for the singlet. Hence $\bar{a}_{\mu}=\left(  a_{\mu
}\right)  ^{\dagger}$ can be satisfied only on singlets or on infinite
dimensional representations of the Lorentz or the SU$\left(  d,1\right)  $
symmetry\footnote{To be more accurate we should distinguish between
fundamental and anti-fundamental representations of SU$\left(  d,1\right)  $
by using differrent indices to label them. For example, we can use undotted
indices $a_{\mu}=\frac{1}{\sqrt{2}}\left(  x_{\mu}+ip_{\mu}\right)  $ to
emphasize that $a_{\mu}$ is in the the fundamental representation and dotted
indices $\bar{a}_{\dot{\mu}}=\frac{1}{\sqrt{2}}\left(  x_{\mu}-ip_{\mu
}\right)  $ to emphasize that $\bar{a}_{\dot{\mu}}$ is in the anti-fundamental
representation. Indices are raised or lowered with the Minkowski metric
$\eta^{\mu\dot{\nu}}$ that has mixed indices, such as $\bar{a}^{\mu}=\eta
^{\mu\dot{\nu}}\bar{a}_{\dot{\nu}},$ and $a^{\dot{\mu}}=\eta^{\dot{\mu}\nu
}a_{\nu}$. Because we will not have much use for it we will forgo this more
accurate notation and use the same type of indices on all creation or
annihilation oscillators. The reader should understand that a lower index on
the operator $\bar{a}$ is really meant to be a dotted index $\bar{a}_{\dot
{\mu}},$ while an upper index on $\bar{a}$ is undotted $\bar{a}^{\mu}.$ The
opposite is true for the operators $a_{\mu},a^{\dot{\mu}}$.}.

In the following we will see that there are different Fock spaces disconnected
from each other, all of which contribute to the full unitary spectrum of $Q$.
These Fock spaces are built with the same oscillators $\bar{a}_{\mu},a_{\nu}$
but are based on three different vacua with different SU$\left(  d,1\right)  $
or SO$\left(  d,1\right)  $ symmetry properties as well as different
space-time geometric properties. This shows that there are some surprising
features of the relativistic harmonic oscillator that are fundamentally
different from the non-relativistic one.

Our aim is to identify the physically acceptable unitary sector of the theory
that contains no ghosts and find ways in which the physical sectors can be
singled out by an appropriate set of constraints.

\section{Symmetric vacuum, non-unitary Fock space}

\label{symmvac}

We will start with the standard approach to the relativistic oscillator Fock
space used by most authors, including string theorists \cite{gsw}. The
corresponding relativistic differential equation $(-\frac{1}{2}\partial^{\mu
}\partial_{\mu}+\frac{1}{2}x^{\mu}x_{\mu})\psi_{\lambda}\left(  x\right)
=\lambda\psi\left(  x\right)  $ in position space, in the purely spacelike
sector, is solved in Appendix A in $1+1$ dimensions. Although the Fock space
approach in this section and the position space approach of Appendix A are in
full agreement, a great deal of complementary insight about the issues
regarding spacetime regions is gained from considering the properties of the
probability amplitude $\psi_{\lambda}\left(  x\right)  $ in position space. So
the reader may benefit from studying the Appendix and comparing it to the Fock
space approach in this section.

What we want to emphasize is that the familiar Fock space approach yields only
part of the quantum states of this relativistic system. After explaining this,
we will discuss a much larger Fock space of quantum states in the following section.

The oscillator approach begins by assuming a normalized \textit{Lorentz
invariant} vacuum state that has \textit{finite positive norm }and is
annihilated by the operators\textit{\ }$a_{\mu}$\textit{\ }%
\begin{equation}
\langle0|0\rangle=1,\;a_{\mu}|0\rangle=0,\;L_{\mu\nu}|0\rangle=0. \label{vac1}%
\end{equation}
The U$\left(  1\right)  $ quantum number or the level number of this state is
zero%
\begin{equation}
J_{0}|0\rangle=\bar{a}\cdot a|0\rangle=0. \label{Jos}%
\end{equation}

A usually unstated property of this vacuum is that it also requires a
spacelike region for $x^{\mu}$ as well as for $p^{\mu}$ since, as a
probability amplitude in position space or momentum space, it has the form%
\begin{equation}
\langle x|0\rangle\sim e^{-x^{2}/2}\text{ and \ }\langle p|0\rangle\sim
e^{-p^{2}/2},\;\;x^{\mu},p^{\mu}\;\text{spacelike.} \label{vac}%
\end{equation}
The minus sign in the exponent follows from satisfying $a_{\mu}|0\rangle=0$ in
position or momentum spaces, namely%
\begin{equation}
a_{\mu}|0\rangle=\frac{1}{\sqrt{2}}\left(  x_{\mu}+ip_{\mu}\right)
|0\rangle=0\leftrightarrow\left\{
\begin{array}
[c]{c}%
\frac{1}{\sqrt{2}}\left(  x_{\mu}+\frac{\partial}{\partial x^{\mu}}\right)
e^{-\frac{1}{2}x\cdot x}=0,\\
\frac{i}{\sqrt{2}}\left(  \frac{\partial}{\partial p^{\mu}}+p_{\mu}\right)
e^{-\frac{1}{2}p\cdot p}=0,
\end{array}
\right\}  .
\end{equation}
Spacelike regions $x\cdot x>0$ and $p\cdot p>0$ are necessary so that the
Gaussian is integrable at infinity%
\begin{equation}
\langle0|0\rangle\sim\int d^{n+1}x~e^{-x^{2}}<\infty,\;\text{or\ \ }%
\langle0|0\rangle\sim\int d^{n+1}p~e^{-p^{2}}<\infty, \label{int}%
\end{equation}
to give a finite norm $\langle0|0\rangle=1.$ Actually these integrals are
infinite as they stand because, unlike the Euclidean analogs in which both
radial and angular integrals are finite, in the present case the
\textquotedblleft angular\textquotedblright\ part contains boost parameters
with an infinite range (see e.g. parametrization in Eq.(\ref{param}) and
Fig.1). For a finite norm this infinity must be divided out (see footnote
(\ref{infinities})).

It is also possible to restrict to a timelike region by starting from another
\textit{Lorentz invariant} \textquotedblleft vacuum\textquotedblright\ state
$|0^{\prime}\rangle$ to construct a different Fock space. This second
alternative is not considered usually. The vacuum $|0^{\prime}\rangle$ is
defined by being annihilated by $\bar{a}_{\mu}$ rather than by $a_{\mu}$
\begin{align*}
\langle0^{\prime}|0^{\prime}\rangle &  =1,\;\;\;\bar{a}_{\mu}|0^{\prime
}\rangle=0,\;\;\;L_{\mu\nu}|0^{\prime}\rangle=0,\\
\;\bar{a}_{\mu}|0^{\prime}\rangle &  =\frac{1}{\sqrt{2}}\left(  x_{\mu
}-ip_{\mu}\right)  |0^{\prime}\rangle=0\leftrightarrow\left\{
\begin{array}
[c]{c}%
\frac{1}{\sqrt{2}}\left(  x_{\mu}-\frac{\partial}{\partial x^{\mu}}\right)
e^{\frac{1}{2}x\cdot x}=0,\\
\frac{i}{\sqrt{2}}\left(  \frac{\partial}{\partial p^{\mu}}-p_{\mu}\right)
e^{\frac{1}{2}p\cdot p}=0.
\end{array}
\right\}
\end{align*}
It corresponds to a normalizable vacuum with $x^{\mu}$ and $p^{\mu}$ in the
timelike region, $x\cdot x<0$ and $p\cdot p<0,$ to be able to normalize
$\langle0^{\prime}|0^{\prime}\rangle=1$
\begin{equation}
\langle x|0^{\prime}\rangle\sim e^{x^{2}/2}\;\;\text{and}\ \;\langle
p|0^{\prime}\rangle\sim e^{p^{2}/2},\;x^{\mu},p^{\mu}\;\text{timelike.}%
\end{equation}
The U$\left(  1\right)  $ quantum number or the level number of this state is
$-\left(  d+1\right)  $%
\begin{equation}
J_{0}|0^{\prime}\rangle=\bar{a}\cdot a|0^{\prime}\rangle=\left[  a\cdot\bar
{a}-\left(  d+1\right)  \right]  |0^{\prime}\rangle=-\left(  d+1\right)
|0^{\prime}\rangle. \label{Jot}%
\end{equation}
so it is clearly distinguishable from the spacelike vacuum.

The Fock space based on the vacuum $|0^{\prime}\rangle$ is not usually
considered because it contains negative norm states for spacelike oscillators,
but by contrast it contains positive norms for timelike oscillators. For
example the 1-particle excitation $a_{\mu}|0^{\prime}\rangle$ has norm
\begin{equation}
\langle0^{\prime}|\bar{a}_{\nu}a_{\mu}|0^{\prime}\rangle=-\eta_{\mu\nu
},\text{~}\left\{
\begin{array}
[c]{l}%
\text{negative for spacelike }\mu,\nu\\
\text{positive for timelike }\mu,\nu
\end{array}
\right.
\end{equation}
However, we will see that the physical states in this Fock space sector
involve always pairs of spacelike and timelike oscillators, such as $a\cdot
a|0^{\prime}\rangle.$ Such paired oscillator states have positive norm. In
this respect, the spacelike or timelike vacua stand at an equal footing. We
will see that while the spacelike vacuum leads to a positive spectrum for $Q,
$ the timelike case leads to a negative spectrum. Whether the negative or
positive spectra are suitable in physical applications depends on the physical
interpretation of the operator $Q=\frac{1}{2}\left(  p\cdot p+x\cdot x\right)
$ in some physical context.

This begins to show that there are several disconnected sectors of Fock spaces
in the spectrum of the relativistic harmonic oscillator. As we will see below
both of these Fock spaces lead to non-unitary vector spaces from which we will
need to fish out a subset of positive norm states. Furthermore, in the next
section, we will discuss a completely different Fock space that is based on a
Lorentz non-invariant vacuum $|\tilde{0}\rangle$ that leads to a completely
unitary infinite dimensional Hilbert space.

In the rest of this section we discuss mainly the Fock space based on the
spacelike vacuum $|0\rangle$ and only give results or make comments about the
very similar Fock space based on the timelike vacuum $|0^{\prime}\rangle$.

In either spacelike or timelike cases, since the vacuum respects the
SO$\left(  d,1\right)  $ symmetry, one should expect to find that all the
states in either Fock space can be classified as irreducible unitary or
non-unitary representations of SO$\left(  d,1\right)  $. Furthermore, the
restriction to a spacelike or timelike region is consistent with an SU$\left(
d,1\right)  $ symmetric vacuum since we can verify that under an infinitesimal
SU$\left(  d,1\right)  $ transformation we obtain
\begin{equation}
J_{\mu\nu}|0\rangle=0,\;J_{\mu\nu}|0^{\prime}\rangle=0, \label{vacsymm}%
\end{equation}
by using the two forms of $J_{\mu\nu}$ given in Eq.(\ref{J}). Hence the Fock
spaces built on these invariant vacua must be classified as complete
irreducible unitary or non-unitary representations not just of SO$\left(
d,1\right)  $ but of SU$\left(  d,1\right)  .$

The total level operator can be written out in more detail as%
\begin{equation}
J_{0}=\bar{a}\cdot a=\left(  -\bar{a}_{0}a_{0}\right)  +\bar{a}_{i}a_{i}%
\end{equation}
Note how the number operator in the timelike direction $\left(  -\bar{a}%
_{0}a_{0}\right)  $ works to give a positive number for the level in the
spacelike Fock space \textit{even when the excitation is in the timelike
direction:\ }$\left(  -\bar{a}_{0}a_{0}\right)  \left[  \bar{a}_{0}%
|0\rangle\right]  =\left(  +1\right)  \left[  \bar{a}_{0}|0\rangle\right]  $%
\begin{equation}
\left(  -\bar{a}_{0}a_{0}\right)  \left[  \bar{a}_{0}|0\rangle\right]
=-\bar{a}_{0}\left[  a_{0},\bar{a}_{0}\right]  |0\rangle=\bar{a}_{0}%
|0\rangle\left(  -1\right)  ^{2}=\left(  +1\right)  \bar{a}_{0}|0\rangle.
\end{equation}
Therefore the total level operator $J_{0}$ on the covariant states $\bar
{a}_{\mu}|0\rangle,$ excited in either the time or space directions $\mu$, has
$J_{0}$ eigenvalue $+1.$

Similarly, the excited states at a general level $J_{0}=n$ in the spacelike
Fock space are constructed by applying $n$ creation operators either in space
or time directions%
\begin{equation}
\bar{a}_{\mu_{1}}\bar{a}_{\mu_{2}}\cdots\bar{a}_{\mu_{n}}|0\rangle
=\text{SU}\left(  d,1\right)  \text{ tensor}\sim\;\overset{n}{\overbrace{%
\begin{tabular}
[c]{|l|l|l|l|l|}\hline
$\mu_{1}$ & $\mu_{2}$ & $\mu_{3}$ & $\cdots$ & $\mu_{n}$\\\hline
\end{tabular}
}}\text{.} \label{yt}%
\end{equation}
This is a symmetric SU$\left(  d,1\right)  $ or U$\left(  d,1\right)  $ tensor
corresponding to a single row Young tableau as indicated. So, this collection
of states at level $J_{0}=n$ form a \textit{finite} dimensional irreducible
representation of SU$\left(  d,1\right)  .$

The above SU$\left(  d,1\right)  $ representation can be reduced into
irreducible representations of SO$\left(  d,1\right)  .$ This is done by
decomposing the symmetric tensor above into a sum of traceless tensors (trace
is defined by contracting with the Minkowski metric $\eta^{\mu\nu}$)
\begin{equation}
\left\{  \left(  \bar{a}_{\mu_{1}}\bar{a}_{\mu_{2}}\cdots\bar{a}_{\mu_{n}%
}-trace\right)  |0\rangle+\cdots\right\}  =\text{SO}\left(  d,1\right)  \text{
traceless tensors.} \label{covariantStates}%
\end{equation}
For example at level $J_{0}=2$ we have one SO$\left(  d,1\right)  $ tensor of
rank $2$ and one of rank zero as listed below
\begin{equation}
\left(  \bar{a}_{\mu_{1}}\bar{a}_{\mu_{2}}-\frac{\eta_{\mu_{1}\mu_{2}}}%
{d+1}\bar{a}\cdot\bar{a}\right)  |0\rangle,\text{ and }\bar{a}\cdot\bar
{a}|0\rangle. \label{rank2}%
\end{equation}
Similarly at level $n$ there are the following irreducible tensors of rank $r
$%
\begin{equation}
r=n,\left(  n-2\right)  ,\left(  n-4\right)  ,\cdots,\left(  0\text{ or
1}\right)  .\text{ } \label{rank}%
\end{equation}
At level $J_{0}=n,$ each traceless tensor of rank $r$ listed in Eq.(\ref{rank}%
) is the basis for a separate \textit{finite} dimensional irreducible
representation of SO$\left(  d,1\right)  .$

All finite representations of non-compact groups, except the singlet, are
non-unitary. Therefore all SU$(d,1)$ or SO$\left(  d,1\right)  $
representations that emerge in this Fock space at all levels $n$, except the
singlets, are non-unitary. Hence at every level $J_{0}=n$ there are many
negative norm states that are unphysical. We have to discuss the types of
constraints that can eliminate the ghosts to obtain a physical theory.

Let us now identify the \textit{negative norm states} which appear among the
SU$\left(  d,1\right)  $ or SO$\left(  d,1\right)  $ states in
Eqs.(\ref{covariantStates},\ref{rank}). These are all the ones that contain an
odd number of \textit{timelike} oscillators. For example, the state $\bar
{a}_{0}|0\rangle$ has negative norm\footnote{The negative norm also implies
that $\langle0|x_{0}x_{0}|0\rangle$ and $\langle0|p_{0}p_{0}|0\rangle$ are
negative as seen from $\langle0|x_{0}x_{0}|0\rangle=\frac{1}{2}\langle
0|\left(  a_{0}+\bar{a}_{0}\right)  \left(  a_{0}+\bar{a}_{0}\right)
|0\rangle=\frac{1}{2}\langle0|a_{0}\bar{a}_{0}|0\rangle=-\frac{1}{2}.$ If
$x_{0}$ were hermitian then $x_{0}x_{0}$ would have to be a positive operator
with positive expectation value. But in this Fock space $x_{0},p_{0}$ are not
hermitian, equivalently $\bar{a}_{0}$ is not the hermitian conjugate of
$a_{0},$ and this is why negative norms arise. \label{hermit}}:
\begin{equation}
\text{norm\ =}\langle0|a_{0}\bar{a}_{0}|0\rangle=\langle0|\left[  a_{0}%
,\bar{a}_{0}\right]  |0\rangle=\left(  -1\right)  \langle0|0\rangle=-1.
\end{equation}
The states at a fixed level $n$ that have an even number of $\bar{a}_{0}$'s
and any number of spacelike oscillators, such as $\left(  \bar{a}_{0}\right)
^{m}\left(  \bar{a}_{i_{1}}\bar{a}_{i_{2}}\cdots\bar{a}_{i_{n-m}}\right)
|0\rangle,$ have positive norm for every even $m=0,2,4,\cdots,\left(  n\text{
or }n-1\right)  $. A constraint that eliminates all negative norm states in
the spacelike region is to demand a reflection symmetry from every state under
the operation $\bar{a}_{0}\rightarrow-\bar{a}_{0}$ and similarly for
$a_{0}\rightarrow-a_{0}.$ This can be achieved through the operator\footnote{A
similar operator for the timelike region is $S=\exp\left(  i\pi\bar{a}%
_{i}a_{i}\right)  .$} $T=\exp\left(  i\pi\bar{a}_{0}a_{0}\right)  $ which
gives $Ta_{0}T^{-1}=-a_{0}$ and $T\bar{a}_{0}T^{-1}=-\bar{a}_{0},$ and the
boost generator changes sign $TL^{0i}T^{-1}=-L^{0i}.$  Therefore a ghost free
spectrum is obtained by demanding the following constraint%
\begin{equation}
T|\phi\rangle=\left(  +1\right)  |\phi\rangle,\text{ }\Leftrightarrow
\;\left\{
\begin{array}
[c]{l}%
\text{ghost free, unitary subset of states,}\\
\text{but not SO}\left(  d,1\right)  \text{ covariant.}%
\end{array}
\right\}  \label{T}%
\end{equation}
However, such states by themselves break the Lorentz symmetry since they
cannot make up complete irreducible representations of SO$\left(  d,1\right)
$ for any non-zero $n$. In the absence of this constraint, in any
\textit{finite} dimensional representation of SO$\left(  d,1\right)  ,$ other
than the singlet, there will always be states with an odd number of timelike
oscillators. For example at level $2$ the irreducible tensor in
Eq.(\ref{rank2}) contains the negative norm states
\begin{equation}
\bar{a}_{0}\bar{a}_{i}|0\rangle.
\end{equation}
Therefore, to eliminate the negative norm states all finite representations of
SO$\left(  d,1\right)  $ must be discarded by some consistent set of
constraints. This leaves only the SO$\left(  d,1\right)  $
singlets\footnote{This is in the case of a single oscillator, as in the
current simplified problem. If there are additional degrees of freedom then
one can find constraints that lead to more interesting ghost-free solutions.
For example, in string theory, with an infinite number of oscillators, the
Virasoro constraints eliminate ghosts while allowing non-singlets of
SO$\left(  d,1\right)  .$ \label{moredofs}} at each even level $J_{0}=2k$%
\begin{equation}
\left(  \bar{a}\cdot\bar{a}\right)  ^{k}|0\rangle,\;k=0,1,2,3,\cdots\text{
positive norm}\;\leftrightarrow\;\text{no ghosts.}%
\end{equation}
The eigenvalue of $Q$ on these states is $\lambda=2k+\frac{d+1}{2}$
\begin{equation}
Q\left[  \left(  \bar{a}\cdot\bar{a}\right)  ^{k}|0\rangle\right]  =\left[
\left(  \bar{a}\cdot\bar{a}\right)  ^{k}|0\rangle\right]  \left(
2k+\frac{d+1}{2}\right)  .\label{singlets}%
\end{equation}
These states have positive norms since $\bar{a}\cdot\bar{a}=-\bar{a}_{0}%
\bar{a}_{0}+\bar{a}_{i}\bar{a}_{i}$ insures that every term in $\left(
-\bar{a}_{0}\bar{a}_{0}+\bar{a}_{i}\bar{a}_{i}\right)  ^{k}$ contains only an
even number of $\bar{a}_{0}$'s. All the SO$\left(  d,1\right)  $ generators
$L^{\mu\nu}$ in Eq.(\ref{L}) annihilate these states since $\left[  L^{\mu\nu
},\bar{a}\cdot\bar{a}\right]  =0$ gives%
\begin{equation}
L^{\mu\nu}\left[  \left(  \bar{a}\cdot\bar{a}\right)  ^{k}|0\rangle\right]
=\left(  \bar{a}\cdot\bar{a}\right)  ^{k}L^{\mu\nu}|0\rangle=0,\text{ Lorentz
singlets.}%
\end{equation}
So, if the Fock space is restricted to the Lorentz invariant subset, then
there are no ghosts.

The position space probability amplitude for these states is determined as%
\[
\psi_{k}^{\left(  +\right)  }\left(  x\right)  \sim\langle x|\left(  \bar
{a}\cdot\bar{a}\right)  ^{k}|0\rangle=\left[  \frac{1}{2}\left(
x-\partial\right)  \cdot\left(  x-\partial\right)  \right]  ^{k}e^{-\frac
{1}{2}x^{\mu}x_{\mu}},\;\text{spacelike }x^{\mu}.
\]
where $\alpha_{k},\tilde{\alpha}_{k}$ are appropriate normalization constants.
For example, for $k=1$ it becomes%
\begin{equation}
\psi_{1}^{\left(  +\right)  }\left(  x\right)  \sim\left(  2x^{2}-\left(
d+1\right)  \right)  e^{-\frac{1}{2}x^{2}}.
\end{equation}
More generally this gives the generalized Laguerre polynomial $L_{k}%
^{\frac{d-1}{2}}\left(  x^{2}\right)  $ with argument $x^{2}$ multiplying the
Gaussian $e^{-\frac{1}{2}x^{\mu}x_{\mu}}.$
\[
\psi_{k}^{\left(  +\right)  }\left(  x\right)  =\alpha_{k}e^{-\frac{1}%
{2}x\cdot x}\sum_{m=0}^{k}\left(  -1\right)  ^{m}\binom{k+\frac{d-1}{2}}%
{k-m}\frac{\left(  x\cdot x\right)  ^{m}}{m!}=\alpha_{k}e^{-\frac{1}{2}x^{2}%
}L_{k}^{\frac{d-1}{2}}\left(  x^{2}\right)  ,
\]
where $\alpha_{k}$ is an overall constant. It can be checked that this
$\psi_{k}^{\left(  +\right)  }\left(  x\right)  $ is indeed a solution of the
relativistic differential equation in $d$+1 dimensions, with the specified
eigenvalue for every positive integer $k$
\begin{equation}
\frac{1}{2}\left[  -\partial^{\mu}\partial_{\mu}+x^{\mu}x_{\mu}\right]
\psi_{k}^{\left(  +\right)  }\left(  x\right)  =\left(  2k+\frac{d+1}%
{2}\right)  \psi_{k}^{\left(  +\right)  }\left(  x\right)  ,\;k=0,1,2,\cdots
\end{equation}
Furthermore, these wavefunctions clearly have positive norm $\int
d^{d+1}x\left\vert \psi_{k}^{\left(  +\right)  }\left(  x\right)  \right\vert
^{2}\;$for all $k.$ We see that according to the symmetry criteria, and
unitarity, only these states are admissible as quantum states in the spacelike
Fock space\footnote{Recall the infinite integrals mentioned following
Eq.(\ref{int}). These resurface again in the norm above. For example, in the
simplified case in Eq.(\ref{squarenorm}) the delta function normalization
$\delta\left(  m^{\prime}-m\right)  $ blows up for $m^{\prime}=m.$ This will
be a common infinite factor for all Lorentz invariant wavefunctions. The
infinity can be avoided by redefining norm by simply not integrating over the
extra boost parameters, since those parameters do not appear in the Lorentz
invariant wavefunctions. If such a redefinition is not adapted, the infinities
may be an argument to discard all of the Lorentz invariant states $\psi
_{k}^{\pm}\left(  x\right)  .$ By comparison note that the unitary states
based on the Lorentz non-invariant vacuum $|\tilde{0}\rangle$ discussed in
section (\ref{unitaryF}) have no infinities in their norms. \label{infinities}%
}.

Similarly, there is another set of SU$\left(  d,1\right)  $ singlet states
$\left(  a\cdot a\right)  ^{k}|0^{\prime}\rangle$ in the timelike Fock space
given by substituting $a_{\mu}$ instead of $\bar{a}_{\mu}$ and using
$|0^{\prime}\rangle$ instead of $|0\rangle$
\begin{gather}
J_{\mu\nu}\left[  \left(  a\cdot a\right)  ^{k}|0^{\prime}\rangle\right]
=0,\\
Q\left[  \left(  a\cdot a\right)  ^{k}|0^{\prime}\rangle\right]  =-\left(
2k+\frac{d+1}{2}\right)  \left[  \left(  a\cdot a\right)  ^{k}|0^{\prime
}\rangle\right]  ,\;\label{minus}\\
\;\psi_{k}^{\left(  -\right)  }\left(  x\right)  =\tilde{\beta}_{k}\left[
\frac{1}{2}\left(  x+\partial\right)  \cdot\left(  x+\partial\right)  \right]
^{k}e^{\frac{1}{2}x^{\mu}x_{\mu}}\sim\langle x|\left(  a\cdot a\right)
^{k}|0^{\prime}\rangle,\;\text{timelike }x^{\mu}.
\end{gather}
The $\psi_{k}^{\left(  -\right)  }\left(  x\right)  $ are related to the
$\psi_{k}^{\left(  +\right)  }\left(  x\right)  $ by an analytic continuation
of $x^{2}\rightarrow-x^{2}$ from the spacelike to the timelike region, so they
can also be expressed in terms of the Laguerre polynomials%
\[
\psi_{k}^{\left(  -\right)  }\left(  x\right)  =\gamma_{k}e^{\frac{1}{2}x\cdot
x}\sum_{m=0}^{k}\binom{k+\frac{d-1}{2}}{k-m}\frac{\left(  x\cdot x\right)
^{m}}{m!}=\gamma_{k}e^{\frac{1}{2}x^{2}}L_{k}^{\frac{d-1}{2}}\left(
-x^{2}\right)  ,
\]
However, it must be emphasized that, as computed\footnote{This follows from
the form of $Q=a\cdot\bar{a}-\frac{d+1}{2}$ given in Eq.(\ref{H}), and from
the fact that $\left[  a\cdot\bar{a},\left(  a\cdot a\right)  \right]
=-2\left(  a\cdot a\right)  .$} in Eq.(\ref{minus}), the $\psi_{k}^{\left(
-\right)  }\left(  x\right)  $ have the opposite signs for the eigenvalues of
$Q$ as compared to the $\psi_{k}^{\left(  +\right)  }\left(  x\right)  .$

All of these $\psi_{k}^{\left(  \pm\right)  }\left(  x\right)  $ are
SO$\left(  d,1\right)  $ invariants, but what are their SU$\left(  d,1\right)
$ properties? The SU$\left(  d,1\right)  $ symmetry of $Q$ and of the vacuum
exhibited in Eqs.(\ref{J},\ref{vacsymm}) require that the spectrum be
classified as complete SU$\left(  d,1\right)  $ multiplets. Which SU$\left(
n,1\right)  $ multiplets do these states correspond to? If we apply an
infinitesimal SU$\left(  d,1\right)  $ transformation on the SO$\left(
d,1\right)  $ singlets, we find
\begin{equation}
J_{\mu\nu}\left[  \left(  \bar{a}\cdot\bar{a}\right)  ^{k}|0\rangle\right]
=2k\left(  \bar{a}_{\mu}\bar{a}_{\nu}-\frac{\eta_{\mu\nu}}{d+1}\left(  \bar
{a}\cdot\bar{a}\right)  \right)  \left(  \bar{a}\cdot\bar{a}\right)
^{k-1}|0\rangle.
\end{equation}
We see on the right hand side that, except for the case of $k=0,$ we generate
inadmissible negative norm states. This also shows that the states $\left(
\bar{a}\cdot\bar{a}\right)  ^{k}|0\rangle$ with $k\neq0$ are not in a singlet
of SU$(d,1)$ so that they must be part of non-unitary representations of
SU$\left(  d,1\right)  .$ Hence even though the states $\left(  \bar{a}%
\cdot\bar{a}\right)  ^{k}|0\rangle$ are unitary with respect to SO$\left(
d,1\right)  ,$ they are not consistent with an SU$\left(  d,1\right)  $
symmetry-consistent unitary spectrum, except for $k=0$.

What happened to the SU$\left(  d,1\right)  $ symmetry? It got broken by the
boundary conditions of restricting the Fock space inadvertently to a purely
spacelike region (see last paragraph of Appendix A for more insight). If one
wishes to be consistent with SU$\left(  d,1\right)  $ covariance, and also
restrict to the spacelike region, then only the vacuum state can be kept in
the spectrum.

In a broken SU$\left(  d,1\right)  $ scenario all Lorentz singlet states
$\left(  \bar{a}\cdot\bar{a}\right)  ^{k}|0\rangle$ are admissible. Similarly,
in a broken SO$\left(  d,1\right)  $ scenario all states of the form (\ref{T})
with an even number of $a_{0}$'s can be included in the ghost free Hilbert
space. But, in an exact SU$\left(  d,1\right)  $ scenario only the vacuum
state $|0\rangle$ can be included. A similar statement applies to the purely
timelike sector where only the second vacuum state $|0^{\prime}\rangle$ can be included.

We see that, in a SU$\left(  d,1\right)  $ symmetry-consistent spacelike or
timelike Fock spaces, all states other than the vacuum states $|0\rangle
,|0^{\prime}\rangle$ must be thrown away by some consistent set of constraints
since otherwise the theory cannot be both consistent with its SU$\left(
d,1\right)  $ symmetry and also be free of ghosts. One possibility is to
choose the constraint to be $J_{\mu\nu}=0$ but this is too restrictive
because, as we will see, it throws away the big sector of unitary states that
we will discuss in the next section. Less restrictive is a constraint of the
form%
\begin{equation}
\left[  \frac{1}{2}\left(  p^{2}+x^{2}\right)  -\lambda_{0}\right]
=0,\;\text{no ghosts only for\ }\lambda_{0}=\pm\frac{d+1}{2}. \label{cpm}%
\end{equation}
When $\lambda_{0}=\frac{d+1}{2}$ the constraint can be satisfied only by
$|0\rangle$ and when $\lambda_{0}=-\frac{d+1}{2}$ it can be satisfied only by
$|0^{\prime}\rangle.$ For other values of $\lambda_{0}$ that appeared in the
spectrum above, such as $\lambda=\pm\left(  n+\frac{d+1}{2}\right)  ,$ the
constraint allows also negative norm states in non-unitary representations of
SU$\left(  d,1\right)  $ with a Young tableau with $n$ boxes as in
Eq.(\ref{yt}), so only $n=0$ is admissible. We see that the only possible
constraint of this form can only involve $\lambda_{0}=\pm\frac{d+1}{2},$
leading to only one of the possible states: either $|0\rangle$ or $|0^{\prime
}\rangle.$

A constraint of the type (\ref{cpm}) with general $\lambda_{0}$ emerges as a
natural outcome in a worldline theory as a consequence of a gauge symmetry on
the worldline as we will see in detail in section (\ref{world}). That kind of
local symmetry is reasonable because it can be used to eliminate the ghosts
that come from timelike directions, thus guaranteeing a unitary theory.

If $\lambda_{0}$ is in the range $-\frac{d+1}{2}<\lambda<\frac{d+1}{2}$ no
state in the spacelike or timelike sectors can satisfy the constraint
(\ref{cpm}). So, with such a constraint all the states in the purely spacelike
or purely timelike sectors, including $|0\rangle$ and $|0^{\prime}\rangle$
would be excluded.

But in the next section we will find that this type of constraint is satisfied
by many more states beyond those that appeared in the spacelike or timelike
Fock spaces discussed in this section. There is a large sector of positive
norm quantum states that cannot be built by starting from the conventional
Lorentz invariant vacuum states $|0\rangle,|0^{\prime}\rangle,$ and those
additional states are compatible with the SU$\left(  d,1\right)  $ symmetry,
not as singlets, but as infinite dimensional unitary representations whose
Casimir eigenvalues are determined by $\lambda_{0}$.

\section{Unitary Fock space, non symmetric vacuum}

\label{unitaryF}

We will now take a different approach to solving the eigenvalue problem
$Q\psi_{\lambda}=\lambda\psi_{\lambda}.$ Rather starting with a Lorentz
invariant vacuum state as is usually done, we will consider solving the
differential equation
\begin{equation}
\frac{1}{2}\left(  -\partial^{\mu}\partial_{\mu}+x^{\mu}x_{\mu}\right)
\psi_{\lambda}\left(  x\right)  =\lambda\psi_{\lambda}\left(  x\right)  .
\end{equation}
without paying attention at first to its Lorentz covariance properties
\cite{yukawa}\cite{dirac}\cite{kim1}. We will then clarify the symmetry
properties of the solutions by appealing to the hidden symmetry SU$\left(
d,1\right)  .$

We can obtain solutions by separating it in the $x^{0},\vec{x}$ variables,
\begin{equation}
\frac{1}{2}\left[  \left(  -\vec{\partial}^{2}+\vec{x}^{2}\right)  -\left(
-\partial_{0}^{2}+x_{0}^{2}\right)  \right]  \psi_{\lambda}\left(  \vec
{x},x_{0}\right)  =\lambda\psi\left(  \vec{x},x_{0}\right)  ,
\end{equation}
with a wavefunction of the form%
\begin{equation}
\psi_{\lambda}\left(  \vec{x},x_{0}\right)  =A_{\lambda_{a}}\left(  \vec
{x}\right)  B_{\lambda_{b}}\left(  x_{0}\right)  ,\;\;\lambda=\left(
\lambda_{a}-\lambda_{b}\right)  ,\label{general1}%
\end{equation}
such that%
\begin{equation}
\frac{1}{2}\left(  -\vec{\partial}^{2}+\vec{x}^{2}\right)  A_{\lambda_{a}%
}\left(  \vec{x}\right)  =\lambda_{a}A_{\lambda_{a}}\left(  \vec{x}\right)
,\;\;\;\frac{1}{2}\left(  -\partial_{0}^{2}+x_{0}^{2}\right)  B_{\lambda_{b}%
}\left(  x_{0}\right)  =\lambda_{b}B_{\lambda_{b}}\left(  x_{0}\right)  .
\end{equation}
In a unitary Hilbert space in which $x^{\mu},p^{\mu}$ are all hermitian
operators, both $\lambda_{a}$ and $\lambda_{b}$ must be positive since the
operators $\frac{1}{2}\left(  \vec{p}^{2}+\vec{x}^{2}\right)  $ as well as
$\frac{1}{2}\left(  p_{0}^{2}+x_{0}^{2}\right)  $ are positive. In fact, from
the study of the Euclidean harmonic oscillator in $d$ dimensions and $1$
dimension respectively we already know all possible eigenvalues and
eigenstates\footnote{The wavefunction of an arbitrary excited state of the
$d$-dimensional Euclidean harmonic oscillator at eigenvalue $\lambda=n+d/2,$
and SO$\left(  d\right)  $ orbital angular momentum quantum number $l,$ has
the form
\[
A_{i_{1}i_{2}\cdots i_{l}}^{nl}\left(  \vec{x}\right)  =e^{-\vec{x}^{2}%
/2}\left\vert \vec{x}\right\vert ^{l}L_{n}^{l-1+d/2}\left(  \vec{x}%
^{2}\right)  T_{i_{1}i_{2}\cdots i_{l}}\left(  \hat{x}\right)  .
\]
Here $T_{i_{1}i_{2}\cdots i_{l}}\left(  \hat{x}\right)  $ is the symmetric
\textit{traceless} tensor of rank $l$ constructed from the unit vector
$\hat{x}_{i}\equiv x_{i}/\left\vert \vec{x}\right\vert $ (this is equivalent
to the spherical harmonics in $d=3$ space dimensions). $L_{\alpha}^{\beta
}\left(  z\right)  $ is the generalized Laguerre polynomial with argument
$z=\vec{x}^{2},$ and indices $\alpha=n$ and $\beta=l-1+d/2.$ The quantum
numbers take the following values: The excitation level $n$ is any positive
integer $n=0,1,2,3,\cdots,$ while at fixed $n$ the allowed values of $l$ are
$l=n,\left(  n-2\right)  ,\left(  n-4\right)  ,\cdots,\left(  1\text{ or
}0\right)  .$ \label{dsolution}} for $\left(  \lambda_{a},A_{\lambda_{a}%
}\left(  \vec{x}\right)  \right)  $ and for $\left(  \lambda_{b}%
,B_{\lambda_{b}}\left(  x_{0}\right)  \right)  ,$ where%
\begin{equation}%
\begin{array}
[c]{c}%
\lambda_{a}=n_{a}+\frac{d}{2},\text{ with }n_{a}=0,1,2,3,\cdots,\\
\lambda_{b}=n_{b}+\frac{1}{2},\text{ with }n_{b}=0,1,2,3,\cdots.
\end{array}
\label{general}%
\end{equation}
Furthermore, we know that the wavefunctions take the
form$^{\text{\ref{dsolution}}}$%
\begin{equation}%
\begin{array}
[c]{c}%
A_{\lambda_{a}}\left(  \vec{x}\right)  =e^{-\frac{1}{2}\vec{x}^{2}}%
\times\text{(polynomial of degree }n_{a}\text{ in the variables }%
x_{i}\text{),}\\
B_{\lambda_{a}}\left(  x_{0}\right)  =e^{-\frac{1}{2}x_{0}^{2}}\times
\text{(polynomial of degree }n_{b}\text{ in the variable }x_{0}\text{).}%
\end{array}
\end{equation}
In this basis there is infinite degeneracy for the same eigenvalue of
$Q\rightarrow\lambda$, since eigenstates with different $n_{a},n_{b}$ can lead
to the same eigenvalue $\lambda=\lambda_{a}-\lambda_{b}=n+\frac{d-1}{2}.$ Thus
with both $m,n$ even integers or with both $m,n$ odd integers we can write
\begin{equation}%
\begin{array}
[c]{c}%
n_{a}=\frac{m+n}{2},\;\;n_{b}=\frac{m-n}{2},\\
\text{at fixed }n,\;\text{all }m\geq\left\vert n\right\vert \text{ gives
infinite degeneracy.}%
\end{array}
\end{equation}
All solutions with the same eigenstate $\lambda$ can be constructed from
(infinite) linear combinations of the ones above, but they all must have the
form
\begin{gather}
\psi_{\lambda}\left(  x^{\mu}\right)  =e^{-\frac{1}{2}\left(  \vec{x}%
^{2}+x_{0}^{2}\right)  }\times\text{(polynomials in the variables }x_{\mu
}\text{),}\label{gaussian}\\
\lambda=n+\frac{d-1}{2},\;\text{with\ }n=0,\pm1,\pm2,\pm3,\cdots,\;\nonumber
\end{gather}
It is evident that these solutions have positive norm since the integrals
converge in all spacetime directions and they are positive%
\begin{equation}
\langle\psi_{\lambda}|\psi_{\lambda}\rangle=\int d^{d+1}x\left\vert
\psi_{\lambda}\left(  x^{\mu}\right)  \right\vert ^{2}=1.
\end{equation}
We have at hand definitely a\textit{\ unitary basis}, but what are the Lorentz
symmetry properties of these solutions?

The striking contrast to the solutions in the previous section is that the
exponent $\left(  \vec{x}^{2}+x_{0}^{2}\right)  $ is not Lorentz invariant,
and hence these solutions and the solutions of the previous section are
mutually exclusive\textit{. }They each span different Hilbert spaces and the
spacetime geometric properties are very different. The Lorentz symmetry
properties of the solutions (\ref{gaussian}) are not yet evident.

On the other hand, the operator $Q$ is invariant under SU$\left(  d,1\right)
$ and its Lorentz subgroup SO$\left(  d,1\right)  $, so we must be able to
organize the solutions at each value $\lambda$ in terms of the representations
of SU$\left(  d,1\right)  $ and any of its subgroups. These representations
are automatically unitary since we have already insured that $x^{\mu},p^{\mu
},$ and therefore the Lorentz generators $L_{\mu\nu}=x_{\mu}p_{\nu}-x_{\nu
}p_{\mu},$ are hermitian in this basis. Hence, we must expect infinite
dimensional unitary representations of SO$\left(  d,1\right)  $ and of
SU$\left(  d,1\right)  $ at each $\lambda.$ In fact, this is in agreement with
the infinite degeneracy at each $\lambda$ noted above. There remains to answer
what precisely are these unitary representations, and how to label states with
quantum numbers within the representation?

We now answer this question. We will explain below that at each $\lambda$
there is a \textit{single irreducible unitary representation of SU}$\left(
d,1\right)  $ whose Casimir eigenvalues are completely determined by $\lambda$
and $d.$ We will give the detailed content of this representation in the group
theoretical basis when SU$\left(  d,1\right)  $ is decomposed into SU$\left(
d\right)  \times$U$\left(  1\right)  .$ In this way we will be able to
determine the SU$\left(  d\right)  ,$ and the angular momentum SO$\left(
d\right)  \subset$SU$\left(  d\right)  ,$ quantum numbers of each quantum state.

The starting point is a new vacuum state $|\tilde{0}\rangle$ which is
different than the Lorentz invariant vacuum states $|0\rangle,$ $|0^{\prime
}\rangle$ of the previous section. The new vacuum state is defined as the
state for which the excitation numbers $n_{a},n_{b}$ are both zero. Hence, it
is defined by the following equations%
\begin{equation}
\bar{a}_{0}|\tilde{0}\rangle=a_{i}|\tilde{0}\rangle=0,\;\text{so }\bar{a}%
_{0}\text{ rather than }a_{0}\text{ acts as annihilator.}%
\end{equation}
The position space representation of this state justifies this definition
since the oscillators $a_{\mu},\bar{a}_{\mu}$ defined in Eq.(\ref{abara}) have
the following form in position space and therefore they act on the state
$|\tilde{0}\rangle$ as creators/annihilators as indicated
\begin{gather}
\langle x|\tilde{0}\rangle\sim\exp\left(  -\frac{x_{0}^{2}+\vec{x}^{2}}%
{2}\right) \\
\text{annihilators:\ }\bar{a}_{0}=\frac{1}{\sqrt{2}}\left(  x_{0}%
+\frac{\partial}{\partial x_{0}}\right)  ,\;a_{i}=\frac{1}{\sqrt{2}}\left(
x_{i}+\frac{\partial}{\partial x^{i}}\right) \label{abar0}\\
\text{creators:\ \ }a_{0}=\frac{1}{\sqrt{2}}\left(  x_{0}-\frac{\partial
}{\partial x_{0}}\right)  ,\;\bar{a}_{i}=\frac{1}{\sqrt{2}}\left(  x_{i}%
-\frac{\partial}{\partial x^{i}}\right)
\end{gather}
The extra sign in front of $\frac{\partial}{\partial x_{0}}$ in $a_{0},\bar
{a}_{0}$ is due to lowering the timelike index with the Minkowski metric
$p_{0}=-i\frac{\partial}{\partial x^{0}}=+i\frac{\partial}{\partial x_{0}}$.
Then it is convenient to define the excitation number operators as\footnote{It
may be helpful to define a new notation for the timelike oscillators, $\bar
{a}_{0}\equiv b$ and $a_{0}\equiv\bar{b},$ so that the operators that have the
bar on top, namely $\bar{b},\bar{a}_{i}$ are creation operators. Indeed the
$b,\bar{b}$ satisfy the usual commutation rules with the $+1$ on the right
hand side, $\left[  b,\bar{b}\right]  =\left[  \bar{a}_{0},a_{0}\right]  =+1$
similar to $\left[  a_{i},\bar{a}_{j}\right]  =\delta_{ij}.$ Then $N_{b}%
=\bar{b}b=a_{0}\bar{a}_{0}$ is the familiar excitation number. \label{b}}%
\begin{equation}
\hat{N}_{a}=\bar{a}_{i}a_{i},\;\hat{N}_{b}=a_{0}\bar{a}_{0},
\end{equation}
where the orders of $a_{0}\bar{a}_{0}$ are reversed compared to traditional
notation. The eigenvalues $\left(  n_{a},n_{b}\right)  $ of these operators
vanish on $|\tilde{0}\rangle$%
\begin{equation}
\hat{N}_{a}|\tilde{0}\rangle=0,\;\hat{N}_{b}|\tilde{0}\rangle=0.
\end{equation}
It should be noted that the Lorentz covariant commutation rule in the timelike
direction $\left[  a_{0},\bar{a}_{0}\right]  =-1$ indicates that an excited
state of the form $\left(  a_{0}\right)  ^{n_{b}}|\tilde{0}\rangle$ is
correctly identified as an eigenstate of $\hat{N}_{b}=a_{0}\bar{a}_{0}$ with
eigenvalue $n_{b}$
\begin{equation}
\hat{N}_{b}\left\{  \left(  a_{0}\right)  ^{n_{b}}|\tilde{0}\rangle\right\}
=\left[  a_{0}\bar{a}_{0},\left(  a_{0}\right)  ^{n_{b}}\right]  |\tilde
{0}\rangle=a_{0}\left[  \bar{a}_{0},\left(  a_{0}\right)  ^{n_{b}}\right]
|\tilde{0}\rangle=n_{b}\left\{  \left(  a_{0}\right)  ^{n_{b}}|\tilde
{0}\rangle\right\}  .
\end{equation}
The general state of the form (\ref{general}) with $n_{a},n_{b}$ excitations
has the Fock space representation%
\begin{equation}
|n_{a},n_{b}\rangle=\left(  \bar{a}_{i_{1}}\bar{a}_{i_{2}}\cdots\bar
{a}_{i_{n_{a}}}\right)  \left(  a_{0}\right)  ^{n_{b}}|\tilde{0}\rangle,
\label{general4}%
\end{equation}
where each index $i_{k}$ labels a vector of SO$\left(  d\right)  $ as well as
the fundamental representation of SU$\left(  d\right)  .$

In term of these, the total level operator $J_{0}=\bar{a}_{i}a_{i}-\bar{a}%
_{0}a_{0}$ which we identified in Eq.(\ref{J0}) becomes $J_{0}=\bar{a}%
_{i}a_{i}-a_{0}\bar{a}_{0}-1,$ or
\begin{equation}
J_{0}=\hat{N}_{a}-\hat{N}_{b}-1.
\end{equation}
Therefore the total level of the vacuum state $|\tilde{0}\rangle$ is%
\begin{equation}
J_{0}|\tilde{0}\rangle=\left(  \hat{N}_{a}-\hat{N}_{b}-1\right)  |\tilde
{0}\rangle=\left(  -1\right)  |\tilde{0}\rangle.
\end{equation}
We contrast this $\left(  -1\right)  $ eigenvalue with the $J_{0}$ eigenvalues
of the vacua $|0\rangle,|0^{\prime}\rangle$ which were $0$ and $\left(
-d-1\right)  $ respectively, as shown in Eqs.(\ref{Jos},\ref{Jot}). We also
see that the $Q\rightarrow\lambda$ eigenvalue of the vacuum is $\lambda
=\frac{d-1}{2}$
\begin{equation}
Q|\tilde{0}\rangle=\left(  J_{0}+\frac{d+1}{2}\right)  |\tilde{0}\rangle
=\frac{d-1}{2}|\tilde{0}\rangle.\text{ }%
\end{equation}
Similarly, for the general state $|n_{a},n_{b}\rangle$ we have%
\begin{equation}
J_{0}|n_{a},n_{b}\rangle=\left(  n_{a}-n_{b}-1\right)  |n_{a},n_{b}%
\rangle,\;\;Q|n_{a},n_{b}\rangle=\left(  n_{a}-n_{b}+\frac{d-1}{2}\right)
|n_{a},n_{b}\rangle
\end{equation}
in agreement with Eq.(\ref{gaussian}).

It must now be emphasized that the vacuum state $|\tilde{0}\rangle$ is neither
Lorentz nor SU$\left(  d,1\right)  $ invariant since the Lorentz boost
operators $L_{0i}=i\left(  \bar{a}_{0}a_{i}-\bar{a}_{i}a_{0}\right)  $ or the
SU$\left(  d,1\right)  $ generators $J_{0i}=\bar{a}_{0}a_{i}$ contain two
creation operators. So the vacuum $|\tilde{0}\rangle$ cannot be invariant
under the subset of SO$\left(  d,1\right)  $ or SU$\left(  d,1\right)  $
infinitesimal transformations generated by the operators that contain double
creation
\begin{equation}
L_{0i}|\tilde{0}\rangle\neq0,\;J_{0i}|\tilde{0}\rangle\neq0.
\end{equation}
However, this structure of double creators or double annihilators is tailor
made for the oscillator approach to representation theory for non-compact
groups or supergroups developed in \cite{gunaydin}-\cite{gunaydin2}. Using
those techniques we will classify the states as parts of infinite dimensional
unitary representations as explained below.

First we note that the oscillators $a_{\mu}$ that are in the fundamental
representation of SU$\left(  d,1\right)  $ contain both creation and
annihilation operators (see footnote (\ref{b}) for $a_{0}\equiv\bar{b}$)
\begin{equation}
a_{\mu}=\left(
\begin{array}
[c]{c}%
a_{0}\\
a_{i}%
\end{array}
\right)  =\left(
\begin{array}
[c]{c}%
\bar{b}\\
a_{i}%
\end{array}
\right)  .
\end{equation}
Therefore a general SU$\left(  d,1\right)  $ transformation mixes creation
with annihilation operators. Similarly the anti-fundamental representation
given by $\bar{a}_{\mu}=(%
\begin{array}
[c]{cc}%
\bar{a}_{0} & \bar{a}_{j}%
\end{array}
)=(%
\begin{array}
[c]{cc}%
b & \bar{a}_{j}%
\end{array}
)$ has the same property, and so does the adjoint representation of SU$\left(
d,1\right)  $ which classifies the generators as the traceless product of the
fundamental and anti-fundamental%
\begin{equation}
J_{\mu\nu}=\bar{a}_{\mu}a_{\nu}-\frac{\eta_{\mu\nu}}{d+1}\bar{a}\cdot
a=\left(
\begin{array}
[c]{cc}%
\bar{a}_{0}a_{0}+\frac{\bar{a}\cdot a}{d+1} & \bar{a}_{0}a_{j}\\
\bar{a}_{i}a_{0} & \bar{a}_{i}a_{j}-\frac{\delta_{ij}}{d+1}\bar{a}\cdot a
\end{array}
\right)  \equiv\left(
\begin{array}
[c]{cc}%
J_{00} & J_{0j}\\
J_{i0} & J_{ij}%
\end{array}
\right)  \label{bigJJ}%
\end{equation}
All of these $J_{\mu\nu}$ are symmetries of the operator $Q$ as we noted
earlier. The double annihilation part of $J_{\mu\nu}$ is the upper right
corner $J_{0j}=\bar{a}_{0}a_{j}=ba_{j}$ and the double creation part is the
lower left corner $J_{i0}=\bar{a}_{i}\bar{b}$ of this matrix. Note that the
$d\times d$ matrix $J_{ij}$ has a traceless part $q_{ij}$ while its trace is
related to the remaining generator $J_{00}$ as follows%
\begin{equation}
J_{ij}=q_{ij}+\delta_{ij}\frac{J_{00}}{d},\;\;J_{00}=\frac{q_{0}}{d+1}.
\end{equation}
The generators of the subgroup SU$\left(  d\right)  \times$U$_{q}\left(
1\right)  \times$U$_{J_{0}}\left(  1\right)  \subset$SU$\left(  d,1\right)
\times$U$_{J_{0}}\left(  1\right)  $ are then
\begin{equation}
q_{ij}=\bar{a}_{i}a_{j}-\frac{\delta_{ij}}{d}\hat{N}_{a},\;\hat{q}_{0}=\hat
{N}_{a}+d\left(  \hat{N}_{b}+1\right)  ,\;J_{0}=\hat{N}_{a}-\hat{N}%
_{b}-1.\label{quanta}%
\end{equation}

The general excited state in Eq.(\ref{general4}) $|n_{a},n_{b}\rangle$ can now
be identified by its SU$\left(  d\right)  \times U_{q}\left(  1\right)  \times
U_{J_{0}}\left(  1\right)  $ quantum numbers, by using a Young tableau as
follows%
\begin{align}
|n_{a},n_{b}\rangle &  =\left(  \bar{a}_{i_{1}}\bar{a}_{i_{2}}\cdots\bar
{a}_{i_{n_{a}}}\right)  \left(  a_{0}\right)  ^{n_{b}}|\tilde{0}%
\rangle\label{nanb}\\
&  =[~\overset{n_{a}}{\overbrace{%
\begin{tabular}
[c]{|l|l|l|l|l|}\hline
$i_{1}$ & $i_{2}$ & $i_{3}$ & $\cdots$ & $i_{n_{a}}$\\\hline
\end{tabular}
\ }}~,\;q_{0},~n_{0}]\label{young}\\
q_{0} &  =n_{a}+d\left(  n_{b}+1\right)  ,~n_{0}=n_{a}-n_{b}-1,\;
\end{align}
Note that the eigenvalue $q_{0}$ is a positive integer such that $q_{0}%
-n_{a}=d\left(  n_{b}+1\right)  $ is positive and furthermore it is a non-zero
multiple of $d.$ The Young tableau corresponds to a completely symmetric
SU$\left(  d\right)  $ tensor of rank $n_{a}$ which fully describes the
SU$\left(  d\right)  $ content of the state $|n_{a},n_{b}\rangle$. This tensor
together with the labels $\hat{q}_{0}\rightarrow q_{0}$ and $J_{0}\rightarrow
n_{0},$ or equivalently $Q\rightarrow\lambda=n_{0}+\frac{d+1}{2}=n_{a}%
-n_{b}+\frac{d-1}{2},$ are a complete set of quantum numbers for any
representation of SU$\left(  d,1\right)  \times$U$_{J_{0}}\left(  1\right)  $
that appears in this theory.

The orbital angular momentum $l$ of any state corresponds to its SO$\left(
d\right)  $ representation. The rank $l$ of a traceless symmetric tensor
determines the angular momentum. The completely symmetric tensor of SU$\left(
d\right)  $ in Eq.(\ref{nanb}) is decomposed into traceless symmetric tensors
of rank $l$ as follows%
\begin{equation}
\text{SO}\left(  d\right)  \text{ tensors:\ }l=n_{a},\left(  n_{a}-2\right)
,\left(  n_{a}-4\right)  ,\cdots,\left(  1\text{ or }0\right)  .\label{l}%
\end{equation}
where each state with angular momentum $l$ at levels $n_{b}$ and $n_{a}=l+2r$
is given by%
\begin{equation}
\left(  \bar{a}_{i_{1}}\bar{a}_{i_{2}}\cdots\bar{a}_{i_{l}}-trace\right)
\left(  \bar{a}_{j}\bar{a}_{j}\right)  ^{r}\left(  a_{0}\right)  ^{n_{b}%
}|\tilde{0}\rangle,\label{tracelessSO}%
\end{equation}
Hence the states $|n_{a},n_{b}\rangle$ contain a direct sum of states of the
type (\ref{tracelessSO}) with the angular momenta $l$ specified in Eq.(\ref{l}).

Now we are ready to identify all the states in the same infinite dimensional
representation of SU$\left(  d,1\right)  \times$U$_{J_{0}}\left(  1\right)  .$
For a fixed $J_{0},$ or equivalently a fixed $Q=J_{0}+\frac{d+1}{2}\rightarrow
n+\frac{d-1}{2},$ we must include all the states $|n_{a},n_{b}\rangle$ that
satisfy $n_{a}-n_{b}=n.$ These may be presented as a direct sum of states,
meaning any linear combination of those states
\begin{equation}
\lambda=\frac{d-1}{2}+n:\left\{
\begin{array}
[c]{c}%
\sum_{k=0}^{\infty}\oplus\left(  \bar{a}_{i_{1}}\bar{a}_{i_{2}}\cdots\bar
{a}_{i_{k+n}}\right)  \left(  a_{0}\right)  ^{k}|\tilde{0}\rangle,\text{ if
}n\geq0\\
\sum_{k=0}^{\infty}\oplus\left(  \bar{a}_{i_{1}}\bar{a}_{i_{2}}\cdots\bar
{a}_{i_{k}}\right)  \left(  a_{0}\right)  ^{k+n}|\tilde{0}\rangle,\text{ if
}n\leq0
\end{array}
\right.  \label{towers}%
\end{equation}
More explicitly we give the example of $n=0$ by writing it out%
\begin{equation}
\lambda=\frac{d-1}{2}:\;|\tilde{0}\rangle\oplus\left[  \bar{a}_{i}a_{0}%
|\tilde{0}\rangle\right]  \oplus\left[  \bar{a}_{i}\bar{a}_{j}\left(
a_{0}\right)  ^{2}|\tilde{0}\rangle\right]  \oplus\left[  \bar{a}_{i}\bar
{a}_{j}\bar{a}_{k}\left(  a_{0}\right)  ^{3}|\tilde{0}\rangle\right]
\oplus\cdots,
\end{equation}
and similarly for $n=1,-1$%
\begin{align}
\lambda &  =\frac{d-1}{2}+1:\;\bar{a}_{i}|\tilde{0}\rangle\oplus\left[
\bar{a}_{i}\bar{a}_{j}a_{0}|\tilde{0}\rangle\right]  \oplus\left[  \bar{a}%
_{i}\bar{a}_{j}\bar{a}_{k}\left(  a_{0}\right)  ^{2}|\tilde{0}\rangle\right]
\oplus\cdots,\label{towerm}\\
\lambda &  =\frac{d-1}{2}-1:\;a_{0}|\tilde{0}\rangle\oplus\left[  \bar{a}%
_{i}\left(  a_{0}\right)  ^{2}|\tilde{0}\rangle\right]  \oplus\left[  \bar
{a}_{i}\bar{a}_{j}\left(  a_{0}\right)  ^{3}|\tilde{0}\rangle\right]
\oplus\cdots.
\end{align}
Evidently each distinct value of $\lambda$ completely determines the allowed
$|n_{a},n_{b}\rangle$ and the corresponding SU$\left(  d\right)  \times
$U$\left(  1\right)  \times$U$\left(  1\right)  $ tensors of each infinite
dimensional tower. Note also that for each $\lambda$ there is a single tower.

It is easy to show that each tower at fixed $\lambda$ is an irreducible
representation of SU$\left(  d,1\right)  .$ Under an SU$\left(  d,1\right)  $
group transformation $g=\exp\left(  i\omega^{\mu\nu}J_{\mu\nu}\right)  $
towers with differing eigenvalues $\lambda\neq\lambda^{\prime}$ cannot mix
with each other since $J_{\mu\nu}$ commutes with $Q.$ Hence a single tower
with fixed $\lambda$ is irreducible under the SU$\left(  d,1\right)  $ group
transformation. Furthermore, all the states within each tower mix because the
double creation operators $J_{i0}=\bar{a}_{i}a_{0}=\bar{a}_{i}\bar{b}$ and the
double annihilation operators $J_{0j}=\bar{a}_{0}a_{j}=ba_{j}$ applied
repeatedly mix all the states under the SU$\left(  d,1\right)  $ group
transformation $g=\exp\left(  i\omega^{\mu\nu}J_{\mu\nu}\right)  .$

In fact, all states in a given tower are obtained by applying repeatedly the
double creation SU$\left(  d,1\right)  $ group generators $J_{i0}=\bar{a}%
_{i}a_{0}=\bar{a}_{i}\bar{b}$ on the lowest state%
\begin{equation}
|\text{tower}\rangle_{\lambda}=\left\{  \sum_{k=0}^{\infty}\oplus\left(
J_{i_{1}0}J_{i_{2}0}\cdots J_{i_{k}0}\right)  \right\}  |\text{lowest}%
\rangle_{\lambda}.
\end{equation}
Therefore, only the lowest state in the tower is sufficient to label uniquely
the SU$\left(  d,1\right)  $ content of the entire tower. These unique labels
correspond to the SU$\left(  d\right)  $ Young tableau and the U$_{q}\left(
1\right)  $ charge $\hat{q}_{0}=\hat{N}_{a}+d\left(  \hat{N}_{b}+1\right)  $,
identified in Eq.(\ref{quanta}). These are the appropriate quantum numbers for
the basis SU$\left(  d\right)  \times$U$_{q}\left(  1\right)  \subset
$SU$\left(  d,1\right)  $ at a fixed $\lambda$%
\begin{equation}
|\text{lowest}\rangle_{\lambda}=\left(  \overset{n_{a}}{\overbrace{%
\begin{tabular}
[c]{|l|l|l|l|l|}\hline
$i_{1}$ & $i_{2}$ & $i_{3}$ & $\cdots$ & $i_{n_{a}}$\\\hline
\end{tabular}
\ }},~q_{0}\left(  \lambda\right)  \right)  ,\;q_{0}=n_{a}\left(  d+1\right)
-d\left(  \lambda-\frac{d+1}{2}\right)  .
\end{equation}

We can easily compute the Casimir operators for the irreducible unitary
representations identified above. The quadratic Casimir operator of SU$\left(
d,1\right)  $ is given by%
\begin{equation}
C_{2}=\frac{1}{2}J_{\mu\nu}\eta^{\nu\lambda}J_{\lambda\sigma}\eta^{\sigma\mu
}=\frac{1}{2}\left(  J_{ij}J_{ji}+\left(  J_{00}\right)  ^{2}-J_{i0}%
J_{0i}-J_{0i}J_{i0}\right)  . \label{C2}%
\end{equation}
After inserting the oscillator form of the $J_{\mu\nu}$ given in
Eq.(\ref{bigJJ}), and rearranging the oscillators we find that $C_{2}$ is
rewritten as a function of only the U$_{J_{0}}\left(  1\right)  $ generator
\begin{equation}
C_{2}\left(  \text{SU}\left(  d,1\right)  \right)  =\frac{dJ_{0}}{2}\left(
\frac{J_{0}}{d+1}+1\right)  . \label{C20}%
\end{equation}
Hence $C_{2}$ is diagonal on any state $|n_{a},n_{b}\rangle$
\begin{equation}
C_{2}|n_{a},n_{b}\rangle=\frac{d\left(  n_{a}-n_{b}-1\right)  \left(
n_{a}-n_{b}+d\right)  }{2\left(  d+1\right)  }|n_{a},n_{b}\rangle,
\end{equation}
and it has the same eigenvalue for all the states in the same tower as follows%
\begin{equation}
C_{2}|\text{tower}\rangle_{\lambda}=\frac{1}{2}d\left(  \lambda-\frac{d-1}%
{2}\right)  \left[  \frac{1}{d+1}\left(  \lambda-\frac{d-1}{2}\right)
+1\right]  |\text{tower}\rangle_{\lambda}%
\end{equation}
Similarly, all SU$\left(  d,1\right)  $ Casimir operators $C_{n}\sim Tr\left(
J\right)  ^{n}$ are found to be only a function of $J_{0}$, so all Casimir
eigenvalues are functions of only $\lambda$.

This result on the Casimirs $C_{n}$ confirms that the full SU$\left(
d,1\right)  $ properties of each tower are completely determined by the
eigenvalue of the operator $Q\rightarrow\lambda.$ Indeed, as seen explicitly
in Eqs.(\ref{towers}-\ref{towerm}), all the states in each tower, and their
SU$\left(  d\right)  \times$U$_{q}\left(  1\right)  \times$U$_{J_{0}}\left(
1\right)  $ quantum numbers, are pre-determined by the fixed value of
$\lambda.$

Now that we have determined that each $|$tower$\rangle_{\lambda}$ corresponds
to a \textit{single unitary representation of }SU$\left(  d,1\right)  $, what
can we say about which unitary representations of the Lorentz group SO$\left(
d,1\right)  $ classify the quantum states? In particular which eigenvalues of
the SO$\left(  d,1\right)  $ Casimir $C_{2}=\frac{1}{2}L_{\mu\nu}L^{\mu\nu}$
appear? This is predetermined by the group theoretical branching rules
SU$\left(  d,1\right)  \rightarrow$SO$\left(  d,1\right)  $ as applied to each
representation. From this it is evident that each $|$tower$\rangle_{\lambda}$
of the type (\ref{towers}) can be written as an infinite direct sum of unitary
representations of SO$\left(  d,1\right)  .$%
\begin{equation}
|\text{tower}\rangle_{\lambda}=\sum\oplus|\text{SO}\left(  d,1\right)  \text{
irreps}\rangle_{\lambda}. \label{reduce}%
\end{equation}
It is not easy to see directly in the oscillator formalism precisely which
eigenvalues of $C_{2}\left(  \text{SO}\left(  d,1\right)  \right)  =\frac
{1}{2}L_{\mu\nu}L^{\mu\nu}$ appear in this sum$.$ This is because the natural
Fock basis $|n_{a},n_{b}\rangle$ we used above is labelled by the eigenvalues
of the operators $\hat{N}_{a},\hat{N}_{b}$ which are not simultaneous
observables with this Casimir
\begin{equation}
\left[  \frac{1}{2}L_{\mu\nu}L^{\mu\nu},\hat{N}_{a}\right]  \neq0,\;\;\left[
\frac{1}{2}L_{\mu\nu}L^{\mu\nu},\hat{N}_{b}\right]  \neq0,
\end{equation}
although $\hat{N}_{a}-\hat{N}_{b}$ is. So, we do not expect that the operator
$\frac{1}{2}L_{\mu\nu}L^{\mu\nu}$ would be diagonal in the basis $|n_{a}%
,n_{b}\rangle.$ Indeed if we construct the SO$\left(  d,1\right)  $ Casimir
operator
\begin{align}
C_{2}\left(  \text{SO}\left(  d,1\right)  \right)   &  =\frac{1}{2}L_{\mu\nu
}L^{\mu\nu}=-\frac{1}{2}\left(  J_{\mu\nu}-J_{\nu\mu}\right)  \left(
J^{\mu\nu}-J^{\nu\mu}\right)  ,\\
&  =-\left(  J_{\mu\nu}J^{\mu\nu}\right)  +J_{\mu\nu}J^{\nu\mu}=-\left(
J_{\mu\nu}J^{\mu\nu}\right)  +2C_{2}\left(  \text{SU}\left(  d,1\right)
\right)  ,
\end{align}
we see that the last part $2C_{2}\left(  \text{SU}\left(  d,1\right)  \right)
$ is diagonal on each state of the $|$tower$\rangle_{\lambda}$, but the first
part $J_{\mu\nu}J^{\mu\nu}$ contains double creation and double annihilation
pieces and hence it cannot be diagonal in the basis $|n_{a},n_{b}\rangle.$
However, it is guaranteed that this basis can be rearranged to the form
(\ref{reduce}), as a superposition of unitary representations of the Lorentz
group SO$\left(  d,1\right)  $ with diagonal $\frac{1}{2}L_{\mu\nu}L^{\mu\nu}%
$, simply because at fixed $n$ we have an irreducible representation of
SU$\left(  d,1\right)  .$ When each SO$\left(  d,1\right)  $ representation in
(\ref{reduce}) is branched down to the SO$\left(  d\right)  $ subgroup of
SO$\left(  d,1\right)  $, then the SO$\left(  d\right)  $ quantum numbers must
agree with those given in Eq.(\ref{l}), namely $l=n_{a},\left(  n_{a}%
-2\right)  ,\cdots,\left(  0~\text{or }1\right)  .$ So, we can deduce that
those SO$\left(  d,1\right)  $ representations that contain this set of
angular momenta must enter in expressing $|n_{a},n_{b}\rangle$ in terms of an
SO$\left(  d,1\right)  $ basis.

\section{Unitarity constraints on the full theory}

We have examined above three distinct Fock spaces based on the three vacua
$|0\rangle,|0^{\prime}\rangle,|\tilde{0}\rangle.$ All the states in these Fock
spaces are eigenstates of the same operator $Q.$ After including the unitarity
condition we found all the physically acceptable positive norm states.

In the quantum theory the existence of different sectors is the analog of
different boundary conditions on the solutions of a given differential
equation. We saw that the \textit{unitary} sectors based on $|0\rangle
,|0^{\prime}\rangle$ are all Lorentz invariant and they are distinguished from
each other by being in the spacelike or timelike regions of spacetime. On the
other hand, none of the unitary states $|n_{a},n_{b}\rangle$ or $|$%
towers$\rangle_{\lambda}$ based on the vacuum $|\tilde{0}\rangle$ are Lorentz
singlets, since $C_{2}$ is non-vanishing on any of them. So, the different
sectors may be distinguished on the basis of their Lorentz, SU$\left(
d,1\right)  $ and geometric properties.

In the absence of boundary conditions that naturally emerge for a specific
physical system all sectors are a priori included. How can we insure that
negative norm ghosts will not appear? We saw that although the sector
$|\tilde{0}\rangle$ is free of ghosts, the sectors $|0\rangle,|0^{\prime
}\rangle$ contained them. It is only by imposing unitarity by
\textquotedblleft hand\textquotedblright, or equivalently by requiring Lorentz
singlets (which may be viewed as a boundary condition), that we could
distinguish the positive norm singlets in the sectors $|0\rangle,|0^{\prime
}\rangle.$ However, requiring Lorentz invariants only as boundary conditions
on the solutions of the entire theory eliminates also the $|\tilde{0}\rangle$
sector completely.

A more comprehensive set of constraints is of the form\footnote{In a theory
with more degrees of freedom more general constraints can also be considered,
see footnote (\ref{moredofs}).}%
\begin{equation}
\frac{1}{2}\left(  p^{2}+x^{2}\right)  -\lambda_{0}=0.
\end{equation}
This allows states from all sectors $|0\rangle,|0^{\prime}\rangle,|\tilde
{0}\rangle$ as long as $\lambda_{0}$ is an eigenvalue of $Q=\frac{1}{2}\left(
p^{2}+x^{2}\right)  .$ The possible eigenvalues in each sector were
\begin{align}
|0\rangle &  :\;\lambda=\frac{d+1}{2}+\left(  \text{positive integer}\right)
\\
|0^{\prime}\rangle &  :\;\lambda=-\frac{d+1}{2}-\left(  \text{positive
integer}\right) \\
|\tilde{0}\rangle &  :\;\lambda=\frac{d-1}{2}+\left(  \text{positive or
negative integer}\right)
\end{align}
We argued in Eq.(\ref{cpm}) that the only way to avoid ghosts in the spacelike
or timelike sectors was to choose $\lambda_{0}=\pm\frac{d+1}{2}.$ Such values
of $\lambda_{0}$ include only the vacua $|0\rangle,|0^{\prime}\rangle$
respectively in the spacelike and timelike sectors, and also the infinite
number of states in the $|$tower$\rangle_{\lambda_{0}}$ in the $|\tilde
{0}\rangle$ sector$.$ Moreover, if we choose $\lambda_{0}$ in the range
$\lambda_{0}=0,\pm1,\pm2,\cdots,\pm\frac{d-1}{2}$ we include only the
corresponding towers $|$tower$\rangle_{\lambda_{0}}$ in the $|\tilde{0}%
\rangle$ sector, but no states at all from the spacelike or timelike sectors
based on $|0\rangle,|0^{\prime}\rangle.$

Hence, if the theory is restricted to the following range only\footnote{We
have not discussed at all the possibility of solutions in the spacelike and
timelike sectors that are matched across the lightcone $x^{2}=0$ as outlined
following Eq.(M). It is possible that those are already accounted for in the
$|\tilde{0}\rangle$ sector, but we are not certain if there are additional
ones. If those have $\lambda$'s within the range in Eq.(\ref{range}) they will
be part of the constrained theory.}
\begin{equation}
-\frac{d+1}{2}\leq\frac{1}{2}\left(  p^{2}+x^{2}\right)  \leq\frac{d+1}%
{2},\;\text{unitary range,} \label{range}%
\end{equation}
then it is guaranteed to be a unitary theory without any negative norm ghosts.
If $\frac{1}{2}\left(  p^{2}+x^{2}\right)  $ is taken outside of this range
then there will always be ghosts coming from the sectors $|0\rangle
,|0^{\prime}\rangle.$ For definiteness we list all the quantum states that
satisfy this range%
\begin{align}
\lambda &  =\frac{d+1}{2}:\;|0\rangle\oplus\bar{a}_{i}\sum_{m=0}^{\infty
}\oplus\left(  \bar{a}_{i_{1}}\bar{a}_{i_{2}}\cdots\bar{a}_{i_{m}}\right)
\left(  a_{0}\right)  ^{m}|\tilde{0}\rangle,\label{top}\\
\lambda &  =\frac{d-1}{2}:\;\sum_{m=0}^{\infty}\oplus\left(  \bar{a}_{i_{1}%
}\bar{a}_{i_{2}}\cdots\bar{a}_{i_{m}}\right)  \left(  a_{0}\right)
^{m}|\tilde{0}\rangle,\label{top2}\\
\lambda &  =\frac{d-3}{2}:\;\sum_{m=0}^{\infty}\oplus\left(  \bar{a}_{i_{1}%
}\bar{a}_{i_{2}}\cdots\bar{a}_{i_{m}}\right)  \left(  a_{0}\right)
^{m+1}|\tilde{0}\rangle,\\
&  \vdots\nonumber\\
\lambda &  =-\frac{d-3}{2}:\sum_{m=0}^{\infty}\oplus\left(  \bar{a}_{i_{1}%
}\bar{a}_{i_{2}}\cdots\bar{a}_{i_{m}}\right)  \left(  a_{0}\right)
^{m+d-2}|\tilde{0}\rangle,\\
\lambda &  =-\frac{d-1}{2}:\;\sum_{m=0}^{\infty}\oplus\left(  \bar{a}_{i_{1}%
}\bar{a}_{i_{2}}\cdots\bar{a}_{i_{m}}\right)  \left(  a_{0}\right)
^{m+d-1}|\tilde{0}\rangle,\label{top6}\\
\lambda &  =-\frac{d+1}{2}:\;|0^{\prime}\rangle\oplus\sum_{m=0}^{\infty}%
\oplus\left(  \bar{a}_{i_{1}}\bar{a}_{i_{2}}\cdots\bar{a}_{i_{m}}\right)
\left(  a_{0}\right)  ^{m+d}|\tilde{0}\rangle,\; \label{toplast}%
\end{align}
Note that the cases of $\lambda=\pm\frac{d+1}{2}$ includes the Lorentz
singlets $|0\rangle,|0^{\prime}\rangle,$ but these singlets do not appear for
the other listed values of $\lambda.$ Furthermore note that only for
$\lambda=+\frac{d+1}{2}$ there is an additional $\bar{a}_{i}$ outside of the
sum in Eq.(\ref{top}). This makes $|0\rangle$ evidently orthogonal to the
tower at $\lambda=+\frac{d+1}{2}.$ The lowest state in each case has
SO$\left(  d\right)  $ angular momentum zero $l=0$. Only the case of
$\lambda=-\frac{d+1}{2}$ has \textit{two} zero angular momentum states one of
which is a SU$\left(  d,1\right)  $ singlet while the other is not.

\section{Worldline theory with gauge symmetry}

\label{world}

A theory with constraints is obtained by constructing a gauge invariant
action. Each constraint is the generator of a gauge symmetry. The gauge
symmetry can be used to eliminate degrees of freedom and in particular it can
remove ghosts and render the theory to be unitary.

A constraint of the type%
\begin{equation}
\phi\left(  x,p\right)  =\frac{1}{2}\left(  p^{2}+x^{2}\right)  -\lambda_{0}=0
\label{constr0}%
\end{equation}
is obtained in the following worldline theory%
\begin{equation}
S\left(  \lambda_{0}\right)  =\int d\tau\left(  \dot{x}^{\mu}p_{\mu}-e\left(
\tau\right)  \left[  \frac{1}{2}\left(  p^{2}+x^{2}\right)  -\lambda
_{0}\right]  \right)  \label{action}%
\end{equation}
where $e\left(  \tau\right)  $ is the gauge field that plays a role of a
Lagrange multiplier locally on the worldline at each instant $\tau.$ The gauge
transformations with a local parameter $\Lambda\left(  \tau\right)  $ are%
\begin{equation}
\delta_{\Lambda}x^{\mu}\left(  \tau\right)  =\Lambda\left(  \tau\right)
p^{\mu}\left(  \tau\right)  ,\;\delta_{\Lambda}p^{\mu}\left(  \tau\right)
=-\Lambda\left(  \tau\right)  x^{\mu}\left(  \tau\right)  ,\;\delta_{\Lambda
}e\left(  \tau\right)  =\frac{d}{d\tau}\Lambda\left(  \tau\right)  .
\label{dxdp}%
\end{equation}
The Lagrangian transforms to a total derivative
\begin{equation}
\delta_{\Lambda}S\left(  \lambda_{0}\right)  =\int d\tau\frac{d}{d\tau}\left(
\frac{1}{2}\left(  p^{2}-x^{2}\right)  \Lambda\left(  \tau\right)
-\lambda_{0}\Lambda\left(  \tau\right)  \right)  \rightarrow0. \label{invS}%
\end{equation}
which can be dropped in the variation of the action (note $p^{2}-x^{2},$ not
$p^{2}+x^{2}$). Hence this action has a local gauge symmetry $\delta_{\Lambda
}S=0$.

One consequence of the gauge symmetry is to impose constraint (\ref{constr0})
as the equation of motion for the gauge field
\begin{equation}
0=\frac{\partial S}{\partial e\left(  \tau\right)  }=\phi\left(  x,p\right)
=\frac{1}{2}\left(  p^{2}+x^{2}\right)  -\lambda_{0}. \label{phi}%
\end{equation}
The generator of the gauge transformations is $\phi\left(  x,p\right)  .$
Saying that $\phi\left(  x,p\right)  $ vanishes is equivalent to saying that
the generator of gauge transformations vanishes, meaning that the sector that
satisfies it must be gauge invariant.

There are various ways to quantize the theory defined by the $S\left(
\lambda_{0}\right)  $ above. The first approach is covariant quantization in
which we work with the quantum rules $\left[  x_{\mu},p_{\nu}\right]
=i\eta_{\mu\nu},$ in an enlarged Hilbert space that includes all the degrees
freedom, including the redundant gauge degrees of freedom that are part of
$x^{\mu},p^{\mu}.$ Then among the quantum states in this enlarged space we
pick the gauge invariant physical states by demanding that they satisfy the
vanishing of the gauge generator%
\begin{equation}
\text{gauge invariants : }\left[  \frac{1}{2}\left(  p^{2}+x^{2}\right)
-\lambda_{0}\right]  |\text{physical}\rangle=0.
\end{equation}
If we follow this approach we see that the gauge invariant states $\langle x|
$physical$\rangle=\psi_{\lambda_{0}}\left(  x\right)  $ are only those that
satisfy the differential equation of the relativistic harmonic oscillator with
a fixed eigenvalue $\lambda_{0}$
\begin{equation}
\left(  -\frac{1}{2}\partial^{\mu}\partial_{\mu}+\frac{1}{2}x^{\mu}x_{\mu
}\right)  \psi_{\lambda_{0}}\left(  x\right)  =\lambda_{0}\psi\left(
x\right)  .
\end{equation}
There is no mention of boundary conditions and therefore we must include all
sectors that solve this constraint. This is the problem we analyzed in the
previous sections. From that analysis we conclude that provided $\lambda_{0} $
is chosen as \textit{one} of the quantized values in the range (\ref{range}),
then the resulting theory $S\left(  \lambda_{0}\right)  $ is guaranteed to be
a ghost free unitary theory.

Outside of this range we expect that ghosts will be present. Therefore
$S\left(  \lambda_{0}\right)  $ with $\lambda_{0}$ fixed to any one of the
values $\lambda_{0}=-\frac{d+1}{2},-\frac{d-1}{2},\cdots,\frac{d-1}{2}%
,\frac{d+1}{2},$ leads to a physically acceptable unitary theory.

A second approach is non-covariant quantization in which we first choose a
gauge and solve the constraint once and for all. The phase space that solves
$\frac{1}{2}\left(  p^{2}+x^{2}\right)  =\lambda_{0}$ is then automatically a
parametrization of the gauge invariant sector. However one must be careful
that there may be more than one sector of phase space that can solve this
equation at the classical level. If we choose a gauge in which the timelike
degree of freedom is eliminated, then the remaining Euclidean degrees of
freedom cannot introduce any negative norm ghosts. The quantum states are then
automatically unitary, but one must check that non-linear expressions are
properly quantum ordered so as to insure that the global symmetries of the
theory have not been violated. Only if the global symmetries are treated
properly - in the present case SU$\left(  d,1\right)  $ and its subgroup
SO$\left(  d,1\right)  $ - can one declare that the theory has been
successfully quantized in the gauge fixed version. In what follows we show how
this is done in the present theory defined by the action $S\left(  \lambda
_{0}\right)  ,$ and how the results agree with the SU$\left(  d,1\right)  $
properties of the covariant quantization approach.

\section{Gauge fixed quantization}

We can choose a gauge that reduces the theory to the purely spacelike harmonic
oscillator. Let us first consider the following canonical transformation from
$\left(  x_{0}\left(  \tau\right)  ,p_{0}\left(  \tau\right)  \right)  $ to
$\left(  t\left(  \tau\right)  ,H\left(  \tau\right)  \right)  $ at the
classical level (i.e. quantum ordering ignored)
\begin{equation}
x_{0}\left(  \tau\right)  =\sqrt{2H\left(  \tau\right)  +2c}~\sin\left(
t\left(  \tau\right)  \right)  ,\;p_{0}\left(  \tau\right)  =\sqrt{2H\left(
\tau\right)  +2c}~\cos\left(  t\left(  \tau\right)  \right)  , \label{canon}%
\end{equation}
where $c$ is some constant that will be fixed later. This covers the entire
$\left(  x_{0},p_{0}\right)  $ plane if $H\left(  \tau\right)  +c\geq0.$ The
new set $\left(  t,H\right)  $ is canonical as can be seen by computing the
corresponding term in the Lagrangian
\[
-\dot{x}_{0}p_{0}=-\dot{t}H+\text{ total derivatives.}%
\]
The total derivatives can be dropped since they are irrelevant in the action.
The Lagrangian in Eq.(\ref{action}) takes the form%
\begin{equation}
L=-\dot{t}H+\dot{x}^{i}p_{i}-e\left[  \frac{1}{2}\left(  \vec{p}^{2}+\vec
{x}^{2}\right)  -H-c-\lambda_{0}\right]  ,
\end{equation}
which shows that the constraint $\phi\left(  x,p\right)  $ that vanishes in
the physical sector now has taken the form%
\begin{equation}
\phi\left(  x,p\right)  =\frac{1}{2}\left(  \vec{p}^{2}+\vec{x}^{2}\right)
-H-c-\lambda_{0}=0.
\end{equation}
Next we choose the gauge
\begin{equation}
t\left(  \tau\right)  =\tau,
\end{equation}
and solve the constraint $\phi\left(  x,p\right)  =0$ to determine the
canonical conjugate of the gauge fixed $t,$ namely $H\left(  \tau\right)  $%
\begin{equation}
H=\frac{1}{2}\left(  \vec{p}^{2}+\vec{x}^{2}\right)  -c-\lambda_{0}.
\label{Hc}%
\end{equation}
The gauge fixed form of the action $S\left(  \lambda_{0}\right)  $ above
describes precisely the spacelike non-relativistic harmonic oscillator after
using $\dot{t}=1$%
\begin{equation}
S_{fixed}\left(  \lambda_{0}\right)  =\int d\tau\left(  \partial_{\tau}\vec
{x}\cdot\vec{p}-\left[  \frac{1}{2}\left(  \vec{p}^{2}+\vec{x}^{2}\right)
-c-\lambda_{0}\right]  \right)  . \label{Sfixed}%
\end{equation}
It is possible to fix the constant $c$ in terms of $\lambda_{0},$ but this is
not necessary at this stage because $\left(  -c-\lambda_{0}\right)  $ seems as
an irrelevant constant that may be dropped. We will wait till we compute
SU$\left(  d,1\right)  $ Casimir eigenvalues at the quantum level to learn the
role of $c$ and its relationship to $\lambda_{0}$ when we compare the results
of covariant quantization to those of the gauge fixed quantization.

The quantum states of this non-relativistic harmonic oscillator in $d$
Euclidean dimensions are well known. They are constructed by defining
creation-annihilation operators $a_{i},\bar{a}_{i}$ in the usual way and
applying them on a vacuum $|\hat{0}\rangle$ that diagonalizes this Hamiltonian%
\begin{equation}
a_{i}|\hat{0}\rangle=0,\;\;\langle\vec{x}|\hat{0}\rangle\sim\exp\left(
-\frac{1}{2}\vec{x}^{2}\right)  .
\end{equation}
The general quantum state is a superposition of the following states that make
up a tower
\begin{align}
|\text{tower}\rangle_{\lambda_{0}}  &  =\sum_{n_{a}=0}^{\infty}\oplus
|n_{a}\rangle=\sum_{n_{a}=0}^{\infty}\oplus\left(  \bar{a}_{i_{1}}\bar
{a}_{i_{2}}\cdots\bar{a}_{i_{n_{a}}}\right)  |\hat{0}\rangle
\;\label{statesfixed}\\
&  \sim~\sum_{n_{a}=0}^{\infty}\oplus\overset{n_{a}}{\overbrace{%
\begin{tabular}
[c]{|l|l|l|l|l|}\hline
$i_{1}$ & $i_{2}$ & $i_{3}$ & $\cdots$ & $i_{n_{a}}$\\\hline
\end{tabular}
}}~.
\end{align}

We compare this spectrum to the towers listed in Eqs.(\ref{top2}-\ref{top6}).
From the comparison we see that the gauge fixed version reproduces the
spectrum of the covariant quantum theory for the action $S\left(  \lambda
_{0}\right)  $ at fixed values of $\lambda_{0},$ provided $\lambda_{0}$ is
fixed to one of the values
\begin{equation}
\lambda_{0}=\frac{d+1}{2},\frac{d-1}{2},\frac{d-3}{2},\cdots,-\frac{d-3}%
{2},-\frac{d-1}{2}, \label{fixedLam}%
\end{equation}
but not the value $\lambda_{0}=-\frac{d+1}{2},$ since in that last case there
is an additional state $|0^{\prime}\rangle$ in Eq.(\ref{toplast}) which does
not show up in Eq.(\ref{statesfixed}).

As we will see below, the gauge fixed version (\ref{statesfixed}) reproduces
the subtlety that for $\lambda_{0}=\frac{d+1}{2}$ there is a Lorentz invariant
state $|0\rangle$ as listed in Eq.(\ref{top}). That is, at $\lambda_{0}%
=\frac{d+1}{2}$ the tower in (\ref{statesfixed}) is actually split into two
representations of SU$\left(  d,1\right)  $. But the gauge fixed version could
not reproduce the other Lorentz invariant state $|0^{\prime}\rangle$ at
$\lambda_{0}=-\frac{d+1}{2}$ in Eqs.(\ref{toplast}). Similarly, the unitary
sector $|n_{a},n_{b}\rangle$ for all $n_{b}<n_{a}$ that appears in covariant
quantization is entirely missed in the fixed gauge. By contrast all the states
$|n_{a},n_{b}\rangle$ for $n_{b}\geq n_{a},$ are recovered in the gauge fixed
version (\ref{statesfixed}) even those beyond the list in (\ref{fixedLam}).

The discrepancy between covariant quantization and gauge fixed quantization is
attributable to an assumption made inadvertently when making the gauge choice.
Namely the canonical transformation (\ref{canon}) is valid only when$\sqrt
{H+c}$ is real. After using Eq.(\ref{Hc}), we see that the reality condition
requires%
\begin{equation}
0\leq H+c=\frac{1}{2}\left(  \vec{p}^{2}+\vec{x}^{2}\right)  -\lambda
_{0}\text{ .}%
\end{equation}
Hence, in the present gauge we have evidently limited ourselves to the quantum
states that satisfy $\lambda_{0}\leq\frac{1}{2}\left(  \vec{p}^{2}+\vec{x}%
^{2}\right)  $. This explains why the gauge fixed version of the theory
defined by $S_{fixed}\left(  \lambda_{0}\right)  $ can be related to the
covariant theory $S\left(  \lambda_{0}\right)  $ only under this condition,
and does not necessarily cover all the gauge invariant sectors of the theory
defined by $S\left(  \lambda_{0}\right)  $ (for a similar example in string
theory, see footnote \ref{foldss}). This is consistent with the fact that the
gauge fixed version could not reproduce all the unitary sectors with
$\lambda\geq$ $\frac{d+1}{2}$. In the guaranteed unitary range $-\frac{d+1}%
{2}\leq\lambda_{0}\leq\frac{d+1}{2},$ all the states except the Lorentz
invariant state $|0^{\prime}\rangle$ at $\lambda_{0}=-\frac{d+1}{2}$ are
recovered. The missing state $|0^{\prime}\rangle$ should be recoverable by
exploring other gauge choices, but we will not pursue this more careful gauge
fixing in this paper.

\section{SU$\left(  d,1\right)  $ and SO$\left(  d,1\right)  $ symmetry in
gauge fixed theory}

We now discuss the unitary representations of the global symmetry SU$\left(
d,1\right)  $ and SO$\left(  d,1\right)  $ in the gauge fixed version, paying
attention to quantum ordering of operators. In particular, we want to show
that the gauge fixed version agrees with the covariant version when we compute
eigenvalues of the Casimir operator $C_{2}\left(  \text{SU}\left(  d,1\right)
\right)  $.

In the \textit{gauge fixed} version, the timelike oscillator $\bar{a}%
_{0}=\frac{1}{\sqrt{2}}\left(  x_{0}-ip_{0}\right)  $ is computed in terms of
the spacelike oscillators $a_{i},\bar{a}_{i}$ after inserting the canonical
transformation (\ref{canon}) and the gauge $t\left(  \tau\right)  =\tau.$ At
the classical level this takes the form%
\begin{equation}
\bar{a}_{0}\left(  \tau\right)  =ie^{i\tau}\sqrt{H+c}=ie^{i\tau}\sqrt{\bar
{a}_{i}\left(  \tau\right)  a_{i}\left(  \tau\right)  +c}.
\end{equation}
At the quantum level one must address operator ordering ambiguities. Since $c
$ has not been fixed so far, we absorb all such ambiguities into $c$ and
define the quantum version of $a_{0}$ with the orders of $\bar{a}_{i}a_{i}$ as
given above. We can now compute the generator of U$_{J_{0}}\left(  1\right)  $
at the quantum level in the gauge fixed version and find the constant value
$J_{0}=-c$%
\begin{equation}
J_{0}=\bar{a}\cdot a=-\bar{a}_{0}a_{0}+\bar{a}_{i}a_{i}=-c. \label{J00}%
\end{equation}
Recall that in the covariant version $Q=J_{0}+\frac{d+1}{2},$ so when
$Q,J_{0}$ are fixed to $Q=\lambda_{0}$ and $J_{0}=-c,$ we determine $c$ as
\begin{equation}
c=\frac{d+1}{2}-\lambda_{0}.
\end{equation}
We see that $c$ is positive only if $\lambda_{0}\leq$ $\frac{d+1}{2}.$ This is
necessary since the square root $\sqrt{\bar{a}_{i}a_{i}+c}$ was defined for
all eigenvalues of the operator $\bar{a}_{i}a_{i}$ only if $c$ is positive
$c\geq0.$

The generators of SU$\left(  d,1\right)  $ can now be computed in the gauge
fixed version by inserting the gauge fixed form of $a_{0}$ and $\bar{a}_{0}$
into the expression of $J_{\mu\nu}$ given in Eq.(\ref{J})
\begin{align}
J_{00}  &  =\hat{N}_{a}+\frac{cd}{d+1},\;J_{ij}=\bar{a}_{i}a_{j}+\frac{c}%
{d+1}\delta_{ij}\label{Jg1}\\
J_{0i}  &  =ie^{i\tau}(\hat{N}_{a}+c)^{\frac{1}{2}}a_{i},\;\;J_{i0}%
=-ie^{-i\tau}\bar{a}_{i}(\hat{N}_{a}+c)^{\frac{1}{2}} \label{Jg2}%
\end{align}
where $\hat{N}_{a}=\bar{a}_{i}a_{i}$ is the number operator. Note that
$J_{00}=\delta^{ij}J_{ij}$ is not independent as expected from $\eta^{\mu\nu
}J_{\mu\nu}=0.$ The non-linear generators $J_{0i},J_{j0}$ must satisfy the
following commutation rules according to the SU$\left(  d,1\right)  $ algebra
(the commutator is evaluated with all $\bar{a}_{i}\left(  \tau\right)  $ and
$a_{j}\left(  \tau\right)  $ at equal $\tau$)
\begin{equation}
\left[  J_{0i},J_{j0}\right]  =\delta_{ij}J_{00}-\eta_{00}J_{ji}.
\label{commutet}%
\end{equation}
We can check explicitly that this commutator is indeed satisfied for any
constant $c$. The critical point in the calculation is to use the property
$a_{i}\hat{N}_{a}=(\hat{N}_{a}+1)a_{i},$ leading to $a_{i}f(\hat{N}%
_{a})=f(\hat{N}_{a}+1)a_{i}$ for any function of $\hat{N}_{a},$ and similarly
for the hermitian conjugate, $\bar{a}_{i}f(\hat{N}_{a}+1)=f(\hat{N}_{a}%
)\bar{a}_{i}.$ Then we can compute the commutator $\left[  J_{0i}%
,J_{j0}\right]  $ as follows
\begin{align}
\left[  J_{0i},J_{j0}\right]   &  =\left(  (\hat{N}_{a}+c)^{\frac{1}{2}}%
a_{i}\right)  \left(  \bar{a}_{j}(\hat{N}_{a}+c)^{\frac{1}{2}}\right)
-\left(  \bar{a}_{j}(\hat{N}_{a}+c)^{\frac{1}{2}}\right)  \left(  (\hat{N}%
_{a}+c)^{\frac{1}{2}}a_{i}\right) \label{comm}\\
&  =a_{i}(\hat{N}_{a}-1+c)^{\frac{1}{2}}(\hat{N}_{a}-1+c)^{\frac{1}{2}}\bar
{a}_{j}-\bar{a}_{j}(\hat{N}_{a}+c)a_{i}\nonumber\\
&  =a_{i}(\hat{N}_{a}-1+c)\bar{a}_{j}-\bar{a}_{j}(\hat{N}_{a}+c)a_{i}%
\nonumber\\
&  =(\hat{N}_{a}+c)a_{i}\bar{a}_{j}-(\hat{N}_{a}-1+c)\bar{a}_{j}%
a_{i}\nonumber\\
&  =\delta_{ij}(\hat{N}_{a}+c)+\bar{a}_{j}a_{i}\nonumber\\
&  =\delta_{ij}\left(  J_{00}+\frac{c}{d+1}\right)  +\left(  J_{ji}-\frac
{c}{d+1}\delta_{ij}\right) \nonumber\\
&  =\delta_{ij}J_{00}+J_{ji},
\end{align}
in agreement with SU$\left(  d,1\right)  $ as in Eq.(\ref{commutet}). It is
easy to check that the rest of the commutation rules for SU$\left(
d,1\right)  $ are satisfied%
\begin{equation}
\left[  J_{\mu\nu},J_{\lambda\sigma}\right]  =\eta_{\nu\lambda}J_{\mu\sigma
}-\eta_{\mu\sigma}J_{\lambda\nu}.
\end{equation}
Hence we have constructed correctly the SU$\left(  d,1\right)  $ algebra. This
implies that we have successfully quantized the theory $S\left(  \lambda
_{0}\right)  $ in the gauge fixed version.

We can now learn the properties of the SU$\left(  d,1\right)  $ representation
by analyzing the transformation properties of the states. The Young tableaux
in Eq.(\ref{statesfixed}) already inform us about their transformation
properties under the subgroup SU$\left(  d\right)  .$ To learn the
transformation rules under the coset generators $J_{i0},J_{0i}$ we apply these
non-linear forms on the states. We see that $J_{i0},J_{0i}$ create or
annihilate excitations
\begin{align}
J_{i0}|n_{a}\rangle &  =\bar{a}_{i}(\hat{N}_{a}+c)^{\frac{1}{2}}|n_{a}%
\rangle\sim|n_{a}+1\rangle\sqrt{n_{a}+c},\\
\;J_{0i}|n_{a}\rangle &  =(\hat{N}_{a}+c)^{\frac{1}{2}}a_{i}|n_{a}\rangle
\sim|n_{a}-1\rangle\sqrt{n_{a}-1+c},
\end{align}
so they mix all SU$\left(  d\right)  $ Young tableaux with each other for all
values of $n_{a}$. So SU$\left(  d,1\right)  $ transformations connect all
levels $n_{a}$ to each other, thus showing that the SU$\left(  d,1\right)  $
representation is infinite dimensional as long as $c>0.$

When $c=0,$ we see that all operators $J_{\mu\nu}$ in Eqs.(\ref{Jg1}%
,\ref{Jg2}) annihilate the vacuum state
\begin{equation}
\left[  J_{\mu\nu}\right]  _{c=0}|\hat{0}\rangle=0. \label{singlet}%
\end{equation}
Therefore for $c=0$ the vacuum state is SU$\left(  d,1\right)  $ and Lorentz
invariant and we must identify it with the Lorentz invariant state $|0\rangle$
listed in Eq.(\ref{top})
\begin{equation}
\lbrack~|\hat{0}\rangle\text{ in gauge fixed version with }%
c=0]\;\leftrightarrow\;[~|0\rangle\text{ in covariant version}]\text{.}%
\end{equation}
Furthermore, when $c=0,$ all the states starting with $n_{a}=1$ form an
irreducible infinite dimensional representation, so they can be written just
like Eq.(\ref{top})
\begin{equation}
c=0,~\text{or }\lambda_{0}=\frac{d+1}{2}:\;|\hat{0}\rangle\oplus\bar{a}%
_{i}\sum_{m=0}^{\infty}\oplus\left(  \bar{a}_{i_{1}}\bar{a}_{i_{2}}\cdots
\bar{a}_{i_{m}}\right)  |\hat{0}\rangle. \label{nrall}%
\end{equation}

Hence at $\lambda_{0}=\frac{d+1}{2}$ we have identified a SU$\left(
d,1\right)  $ or SO$\left(  d,1\right)  $ singlet, together with an infinite
dimensional unitary representation of SU$\left(  d,1\right)  $ whose lowest
state has angular momentum $l=1.$ For all the other cases of $-\frac{d-1}%
{2}\leq\lambda_{0}\leq\frac{d-1}{2}$ the lowest state has angular momentum
zero $l=0$ but it is not a Lorentz or SU$\left(  d,1\right)  $ singlet. At
$\lambda_{0}=-\frac{d+1}{2},$ according to covariant quantization in
Eq.(\ref{toplast}), we should expect a Lorentz singlet together with another
zero angular momentum state as part of an infinite dimensional representation,
but the Lorentz invariant state $|0^{\prime}\rangle$ is missed in the gauge
fixed version.

It is interesting to compute the Casimir operator $C_{2}\left(  \text{SU}%
\left(  d,1\right)  \right)  $ in the gauge fixed version. To do so we insert
the gauge fixed $J_{\mu\nu}$ of Eq.(\ref{Jg1},\ref{Jg2}) into Eq.(\ref{C2})
and manipulate orders of operators as in Eq.(\ref{comm}). After rearranging
operators we find that $C_{2}$ is just a constant determined by $c$ as
follows
\begin{align}
C_{2}  &  =\frac{1}{2}\left(  J_{ij}J_{ji}+\left(  J_{00}\right)  ^{2}%
-J_{i0}J_{0i}-J_{0i}J_{i0}\right) \\
&  =\left\{
\begin{array}
[c]{c}%
\frac{1}{2}\left(  \bar{a}_{i}a_{j}+\frac{c}{d+1}\delta_{ij}\right)  \left(
\bar{a}_{j}a_{i}+\frac{c}{d+1}\delta_{ij}\right)  +\frac{1}{2}\left(  \hat
{N}_{a}+\frac{cd}{d+1}\right)  ^{2}\\
-\frac{1}{2}\bar{a}_{i}(\hat{N}_{a}+c)^{\frac{1}{2}}(\hat{N}_{a}+c)^{\frac
{1}{2}}a_{i}-\frac{1}{2}(\hat{N}_{a}+c)^{\frac{1}{2}}a_{i}\bar{a}_{i}(\hat
{N}_{a}+c)^{\frac{1}{2}}%
\end{array}
\right\} \\
&  =\frac{\left(  -c\right)  d}{2}\left(  1+\frac{\left(  -c\right)  }%
{d+1}\right)
\end{align}
This is the same result as the covariant approach (\ref{C20}) with $J_{0}$
fixed in the gauge fixed version to $J_{0}=-c,$ consistent with Eq.(\ref{J00}).

It may be interesting to discuss also the SO$\left(  d,1\right)  $ content of
each tower. The hermitian Lorentz generators are%
\begin{equation}
\text{SO}\left(  d,1\right)  :\;L_{\mu\nu}=x_{\mu}p_{\nu}-x_{\nu}p_{\mu
}=-i\left(  J_{\mu\nu}-J_{\nu\mu}\right)
\end{equation}
which take the following explicit forms in terms of oscillators%
\begin{align}
\text{rotation}  &  :\;L_{ij}=-i\left(  \bar{a}_{i}a_{j}-\bar{a}_{j}%
a_{i}\right)  ,\;\\
\text{boost\ \ }  &  :\text{\ }L_{0i}=-i\left(  (\hat{N}_{a}+c)^{\frac{1}{2}%
}a_{i}-\bar{a}_{i}(\hat{N}_{a}+c)^{\frac{1}{2}}\right)
\end{align}
It is emphasized that these operators satisfy the SO$\left(  d,1\right)  $ Lie
algebra
\begin{equation}
\left[  L_{\mu\nu},L_{\lambda\sigma}\right]  =-i\left(  \eta_{\nu\lambda
}L_{\mu\sigma}+\eta_{\mu\sigma}L_{\nu\lambda}-\eta_{\mu\lambda}L_{\nu\sigma
}-\eta_{\nu\sigma}L_{\mu\lambda}\right)  ,
\end{equation}
and in particular the commutator of two boosts gives SO$\left(  d\right)  $
rotations at the quantum level
\begin{equation}
\left[  L_{0i},L_{0j}\right]  =-iL_{ij}=i\eta_{00}L_{ij}.
\end{equation}
This can be checked explicitly for our non-linear $L_{0i}$ by using the same
methods as Eq.(\ref{comm}).

Since the $L_{\mu\nu}$ are hermitian they act in infinite dimensional unitary
representations of the Lorentz group. This implies that each tower of
SU$\left(  d,1\right)  $ at fixed $\lambda_{0}$ splits into an infinite number
of irreducible SO$\left(  d,1\right)  $ towers, the precise content of which
SO$\left(  d,1\right)  $ representations appear depend on the constant $c.$

In this section we exhibited new interesting \textit{non-linear oscillator
representations} of SU$\left(  d,1\right)  $ which should have generalizations
to other non-compact groups. This type of oscillator representation was not
previously considered in \cite{gunaydin}-\cite{gunaydin2}. The new non-linear
expressions for the generators given in Eqs.(\ref{Jg1},\ref{Jg2}) were
obtained by starting from previous oscillator methods and then replacing some
of those oscillators by non-linear expressions in terms of the other
oscillators. The same method was used to find new interesting SU$\left(
2,3\right)  $ symmetry properties based on twistors \cite{twistorspin} that
describe spinning particles in various 1T-physics systems and explain
dualities among them. This non-linear approach to constructing generators and
representations of non-compact groups could be of interest in many
applications in both physics and mathematics.

\section{Non-relativistic oscillator as a relativistic system}

While the focus in this paper was the relativistic harmonic oscillator, we
were led to the non-relativistic case as a consequence of a gauge choice.
Looking at this process in reverse, this shows that the non-relativistic
oscillator provides a non-linear realization of a relativistic system. So the
non-relativistic oscillator must have some hidden relativistic symmetry of its
own. This is possibly a surprising proposition, but it is true as explained below.

In $d$ Euclidean dimensions the non-relativistic oscillator has evident
SO$\left(  d\right)  $ symmetry and also a well known SU$\left(  d\right)  $
hidden symmetry that leaves the Hamiltonian invariant. However the discussion
above suggests that we should seek an even larger hidden symmetry SU$(d,1)$
that includes Lorentz symmetry SO$\left(  d,1\right)  .$

We recall that the generator of the gauge symmetry of the relativistic action
$S\left(  \lambda_{0}\right)  $ is $\phi\left(  x,p\right)  =Q\left(
x,p\right)  -\lambda_{0}$ as in Eq.(\ref{phi}). By using Poisson brackets
$\delta_{\Lambda}A\left(  x,p\right)  =$ $\Lambda\left\{  A\left(  x,p\right)
,\phi\left(  x,p\right)  \right\}  $ the gauge transformation rules for all
observables $A\left(  x,p\right)  $ are obtained. In particular note that the
gauge transformations of $\delta_{\Lambda}x^{\mu}$ and $\delta_{\Lambda}%
p^{\mu}$ in Eq.(\ref{dxdp}) follow in this way. Since the SU$\left(
d,1\right)  $ generators $J_{\mu\nu}$ commute with the SU$\left(  d,1\right)
$ invariant $Q$ as shown in Eq.(\ref{Ud1}), it must have vanishing Poisson
brackets with the gauge generator $\phi\left(  x,p\right)  $ when the
$J_{\mu\nu}$ of Eq.(\ref{J}) is written out in terms of phase space
\begin{equation}
\left\{  J_{\mu\nu}\left(  x,p\right)  ,\phi\left(  x,p\right)  \right\}
=0\;\leftrightarrow\;\delta_{\Lambda}J_{\mu\nu}=0. \label{invJ}%
\end{equation}
Therefore, the $J_{\mu\nu}$ are gauge invariant physical observables.

Since both $S\left(  \lambda_{0}\right)  $ and its global symmetry generators
$J_{\mu\nu}$ are gauge invariants, it must be that their gauge fixed versions
$S_{fixed}\left(  \lambda_{0}\right)  ,J_{\mu\nu}^{fixed}$ also maintain the
same SU$\left(  d,1\right)  $ global symmetry properties. That is, when
written out in terms of the remaining Euclidean degrees of freedom $\vec
{x},\vec{p}$ we must find that $J_{\mu\nu}^{fixed}$ is the generator of
SU$\left(  d,1\right)  $ symmetry of the non-relativistic harmonic oscillator
action%
\begin{equation}
S_{non.rel.}=\int d\tau\left(  \partial_{\tau}\vec{x}\cdot\vec{p}-\frac{1}%
{2}\left(  \vec{p}^{2}+\vec{x}^{2}\right)  \right)  .
\end{equation}
The explicit form of $J_{\mu\nu}^{fixed}\left(  \vec{x},\vec{p},\tau\right)  $
is obtained directly from Eqs.(\ref{Jg1},\ref{Jg2}). If these $J_{\mu\nu
}^{fixed}$ are symmetry generators they must be conserved when the equations
of motion of the non-relativistic oscillator are used
\begin{equation}
\frac{d}{d\tau}J_{\mu\nu}^{fixed}\left(  \vec{x}\left(  \tau\right)  ,\vec
{p}\left(  \tau\right)  ,\tau\right)  =0,\;\text{for\ \ }%
\begin{array}
[c]{l}%
\dot{x}_{i}=p_{i}\\
\dot{p}_{i}=-x_{i}%
\end{array}
\text{ \ or\ }%
\begin{array}
[c]{l}%
\dot{a}_{i}=-ia_{i}\\
\overset{\centerdot}{\bar{a}}_{i}=+i\bar{a}_{i}%
\end{array}
. \label{conserved}%
\end{equation}
Note that $J_{0i}^{fixed},J_{i0}^{fixed}$ depend explicitly on $\tau$ in
addition to the implicit dependence on $\tau$ that comes through $\vec
{x}\left(  \tau\right)  ,\vec{p}\left(  \tau\right)  .$ Indeed this extra
dependence on $\tau$ is essential to show that the $J_{0i}^{fixed}%
,J_{i0}^{fixed}$ are conserved.

Since we have already shown that these $J_{\mu\nu}^{fixed}\left(  \vec{x}%
,\vec{p},\tau\right)  $ close to form the SU$\left(  d,1\right)  $ Lie algebra
at the quantum level at any $\tau$, they also satisfy the same property at the
classical level under Poisson brackets. Using these generators we can define
infinitesimal SU$\left(  d,1\right)  $ transformation laws by using Poisson
brackets at any fixed $\tau$, namely $\delta_{\omega}\vec{x}=\frac{1}{2}%
\omega^{\mu\nu}\left\{  \vec{x},J_{\mu\nu}^{fixed}\left(  \tau\right)
\right\}  $ and $\delta_{\omega}\vec{p}=\frac{1}{2}\omega^{\mu\nu}\left\{
\vec{p},J_{\mu\nu}^{fixed}\left(  \tau\right)  \right\}  .$ More explicitly
the transformation laws at any $\tau$ are
\begin{equation}
\delta_{\omega}\vec{x}\left(  \tau\right)  =\frac{1}{2}\omega^{\mu\nu}%
\frac{\partial J_{\mu\nu}^{fixed}\left(  x,p,\tau\right)  }{\partial\vec{p}%
},\;\;\delta_{\omega}\vec{p}\left(  \tau\right)  =-\frac{1}{2}\omega^{\mu\nu
}\frac{\partial J_{\mu\nu}^{fixed}\left(  x,p,\tau\right)  }{\partial\vec{x}}.
\label{nrSU}%
\end{equation}
The transformations under the SU$\left(  d\right)  \times$U$\left(  1\right)
$ subgroup are familiar hidden symmetry transformations of the
non-relativistic harmonic oscillator. However, the transformations generated
by the classical
\begin{align}
\frac{1}{\sqrt{2}}\left(  J_{i0}^{fixed}+J_{0i}^{fixed}\right)   &
=\sqrt{\frac{1}{2}\left(  \vec{p}^{2}+\vec{x}^{2}\right)  +c}\left(  x_{i}%
\cos\tau-p_{i}\sin\tau\right)  ,\;\label{expl1}\\
\frac{1}{\sqrt{2}i}\left(  J_{i0}^{fixed}-J_{0i}^{fixed}\right)   &
=\sqrt{\frac{1}{2}\left(  \vec{p}^{2}+\vec{x}^{2}\right)  +c}\left(  x_{i}%
\sin\tau+p_{i}\cos\tau\right)  \label{expl2}%
\end{align}
are new non-linear symmetry transformations that were not noted before. It can
now be verified that the non-relativistic harmonic oscillator action above is
indeed invariant under all of the SU$\left(  d,1\right)  $ transformations. It
can be verified that the new transformations give $\delta_{\omega}%
S_{non.rel.}=\int d\tau\frac{d}{d\tau}\left(  \text{stuff}\right)
\rightarrow0,$ where the total derivative can be dropped in the transformation
of the action, thus verifying the expected SU$\left(  d,1\right)  $ global
symmetry. Again the explicit $\tau$ dependence generated by the expressions in
(\ref{expl1},\ref{expl2}) is crucial for this result. A consequence of this
symmetry via Noether's theorem is that the $J_{i0}^{fixed}\pm J_{0i}^{fixed}$
given in Eq.(\ref{expl1},\ref{expl2}) are conserved, as already claimed above
in Eq.(\ref{conserved}).

This hidden symmetry of the non-relativistic harmonic oscillator was not known
before. These transformations leave the \textit{action}, not the Hamiltonian,
invariant. As a consequence of the symmetry all the states of the
non-relativistic harmonic oscillator taken together at all energy levels must
fit into irreducible unitary representations of SU$\left(  d,1\right)  _{c}$
and its Lorentz subgroup SO$\left(  d,1\right)  .$

Note that the parameter $c$ is used to construct the non-linear generators
$J_{0i}\left(  c\right)  $ and $J_{i0}\left(  c\right)  $ in Eq.(\ref{Jg2}),
so the SU$\left(  d,1\right)  _{c}$ transformations are different for every
$c.$ This means different representations of SU$\left(  d,1\right)  $ can be
realized on the same Fock space consisting of all the states in
Eq.(\ref{statesfixed}). They will transform differently as a representation
basis depending on the choice of the parameter $c.$ When $c\neq0,$ all the
states form a single irreducible representation of SU$\left(  d,1\right)  $
with Casimir eigenvalue $C_{2}\left(  \text{SU}\left(  d,1\right)
_{c}\right)  =-\frac{cd}{2}\left(  1-\frac{c}{d+1}\right)  $. The lowest state
of this infinite tower has zero SO$\left(  d\right)  $ orbital angular
momentum $l=0$ since it is the vacuum state $|\hat{0}\rangle$. The branching
of the SU$\left(  d,1\right)  _{c}$ representation into representations of the
Lorentz group SO$\left(  d,1\right)  $ depend on $c,$ so we expect to describe
different relativistic content by using the same non-relativistic harmonic
oscillator degrees of freedom.

The $c=0$ case is special, because then the vacuum state $|\hat{0}\rangle$ of
the non-relativistic harmonic oscillator is a singlet of SU$\left(
d,1\right)  _{0}$ and of SO$\left(  d,1\right)  ,$ so it is a Lorentz
invariant as explained in Eq.(\ref{singlet}). The remaining states at all
energy levels given in Eq.(\ref{nrall}) make up a single irreducible unitary
representation of SU$\left(  d,1\right)  _{0}$ with Casimir $0.$ The lowest
energy state of this $c=0$ infinite tower is $\bar{a}_{i}|\hat{0}\rangle$
which has SO$\left(  d\right)  $ angular momentum $l=1.$ This is clearly
different SO$\left(  d,1\right)  $ content compared to the $c\neq0$ case for
which the lowest state of the irreducible tower had angular momentum $l=0$.

This different SU$\left(  d,1\right)  $ or SO$\left(  d,1\right)  $
rearrangement of the same states for different values of $c$ seems surprising
when viewed from the perspective of the non-relativistic oscillator. However,
when compared to the corresponding $|$towers$\rangle_{\lambda_{0}}$ in
Eqs.(\ref{top}-\ref{toplast}) in covariant quantization, the hidden
information in $c$ about the SO$\left(  d,1\right)  $ properties become
evident. The comparison shows that $c$ corresponds to the various powers of
$a_{0}$ applied on the vacuum $|\tilde{0}\rangle$ to get the lowest state
$\left(  a_{0}\right)  ^{c-1}|\tilde{0}\rangle$ in different towers (for
$c\geq1$). The additional information gained from the Lorentz properties of
$a_{0}$ in covariant quantization explains why the same non-relativistic Fock
space (\ref{statesfixed}) relates to different relativistic SO$\left(
d,1\right)  $ or SU$\left(  d,1\right)  $ content as the value of $c$ changes.

Note that if the starting point were the non-relativistic oscillator, then
there would be no conditions on the value of $c$ for constructing the
SU$\left(  d,1\right)  _{c}$ generators in Eq.(\ref{Jg2}). Of course when $c$
is quantized as indicated before, $c=0,1,2,\cdots,\left(  d+1\right)  ,$ the
non-linear structures $J_{0i},J_{i0}$ correspond to just a gauge fixed sector
of the relativistic oscillator with a unitarity constraint. Other values of
$c$ on the real line seem to describe relativistic systems beyond the oscillator.

Note that $c$ is a Lorentz invariant, therefore in physical applications it
could be related to certain relativistically invariant observables, such as
the mass of a bound state.

Such relativistic properties of the \textit{non-relativistic} oscillator may
lead to further insights.

\section{More revisits?}

We have shed new light on the symmetries and the quantum sectors of the
relativistic harmonic oscillator. Since much of this was not noted before, it
may lead to additional new observations in old or new applications of this
commonly used dynamical system.

Of course, for each physical system there may be various sets of new
constraints not discussed in this paper that would influence the allowed
physical states as noted in footnote (\ref{moredofs}). In particular the
richer structure of the many oscillators in string theory leads to the
Virasoro constraints for removing ghosts rather than those in Eq.(\ref{cpm}).
Whatever the ghost killing constraints may be, it would be of interest to
reanalyze the relevant systems to find out whether the additional Fock spaces
discussed in this paper lead to additional quantum states that may reveal new
physical properties.

This paper is not focused on string theory, but rather on the single
relativistic harmonic oscillator. Our initial aim was to clarify some facts
about the symmetry aspects of the relativistic oscillator that appeared
confusing. The clarification provided here leads us to ask what happens in
string theory? In what follows we provide some brief \textit{preliminary
remarks} on this topic.

Past work in string theory has been carried out by relying on the Fock space
built exclusively from the covariant \textit{spacelike} vacuum $|0\rangle$ of
section (\ref{symmvac}), while being unaware of the other Fock space sectors
with more general geometry discussed in sections (\ref{symmvac},\ref{unitaryF}%
). As is well known from previous study of string theory, although not made
previously explicit, the spacelike sector is completely consistent. Its
results have been reproduced in many approaches, leading to the remarkable
properties of string scattering amplitudes.

The question that arises now is not whether anything was wrong with that
treatment of strings, but whether there might be more physical phenomena in
string theory beyond the usual self consistent spacelike sector, and hence
beyond the Veneziano amplitudes. The question is natural since the
conventional relativistic Fock space used in string theory inadvertently
excludes a huge sector of unitary quantum states for each single mode as
discussed in section (\ref{unitaryF}). As made clear following Eq.(\ref{M}),
the relativistic oscillator actually likes to cross between spacelike and
timelike regions. Such allowed motions of each single mode simply have never
entered the discussion, and therefore there is much room for investigation.

In that connection, it is worth noting that from the earliest period of string
theory there has been indications that the lightcone gauge fails to capture
all of the gauge invariant physics in string theory (see footnote
(\ref{foldss})). A similar phenomenon of missing gauge invariant sectors was
seen in the gauge fixed relativistic oscillator discussed in this paper.
Therefore gauge fixed treatments, while being quite revealing, cannot be
trusted as being complete.

These observations provide new motivation to revisit the covariant
quantization of string theory to see whether the concepts discussed in this
paper play a role. In the standard treatment of string theory each mode is
associated with the spacelike vacuum $|0\rangle$, so the standard overall
string vacuum is $|0,0,0,\cdots\rangle,$ where each $0$ corresponds to a mode.
Is it possible to have string configurations built on more complicated vacua,
such as $|0,\tilde{0},0^{\prime},\cdots\rangle$ etc. where the various modes
could be in various spacetime regions? It is not so easy to answer this
question because of the Virasoro constraints.

The sector with all the modes in the timelike Fock space based on $|0^{\prime
},0^{\prime},0^{\prime},\cdots\rangle,$ abreviated as $|0^{\prime}\rangle,$ is
not difficult to decipher because the analysis is parallel to the usual
treatment. The only change is that in this sector all creation annihilation
operators $\alpha_{n}^{\mu},\alpha_{-n}^{\mu}$ switch roles relative to the
familiar spacelike sector. Then we find that this sector has a lot of serious
problems. The eigenvalues of $Q_{n}=\frac{1}{2}\left(  p_{n}^{2}+n^{2}%
x_{n}^{2}\right)  $ are strictly negative and $L_{0}^{\prime}=p_{0}^{2}%
+\sum_{n}Q_{n}+a,$ which is normal ordered$^{\ref{a}}$ relative to
$|0^{\prime}\rangle,$ has only negative eigenvalues. Hence the Virasoro
constraint $L_{0}^{\prime}=1$ gives only tachyons. The Virasoro constraints
$L_{-n}|\phi^{\prime}\rangle$ with $n>0$ (not $L_{n}$) can be satisfied by
using the same arguments as \cite{goddard}-\cite{thorn} but switching
$\alpha_{n}^{\mu}$ with $\alpha_{-n}^{\mu}$ at every step. However, the
solutions still have ghosts at every mass level because the oscillators
$\alpha_{n}^{i}$ in $d$ space dimensions produce negative norm states (as
opposed to only one time component $\alpha_{-n}^{0}$ in the usual arguments).
Evidently this sector is not acceptable on physical grounds and must be
eliminated with some consistent set of gauge symmetries or other arguments.
The supersymmetric version of string theory may avoid this sector alltogether,
but this needs to be investigated more explicitly.

A more interesting case is the ghost free fully unitary sector based on the
vacuum of type $|\tilde{0},\tilde{0},\tilde{0},\cdots\rangle$ which we
abbreviate as $|\tilde{0}\rangle$. For example the string state $|k,\tilde
{0}\rangle$ has a spacetime configuration of the form (note the relative $+$
sign in $(x_{n0}^{2}+\vec{x}_{n}^{2})$)
\begin{equation}
\psi\left(  X\right)  \sim\langle X|k,\tilde{0}\rangle\sim e^{ik\cdot x_{0}%
}\exp\left(  -\frac{1}{2}\sum_{n=1}^{\infty}n(x_{n0}^{2}+\vec{x}_{n}%
^{2})\right)  ,
\end{equation}
where $x_{n}^{\mu}$ can be in any spacetime region unlike the usual string
field in Eq.(\ref{tachyon}) where $x_{n}^{\mu}$ was strictly spacelike. This
is one of the eigenstates of $L_{0}$. There are now an infinite number of
eigenstates for each eigenvalue of $Q_{n}=\frac{1}{2}\left(  p_{n}^{2}%
+n^{2}x_{n} ^{2}\right)  ,$ as explained in section (\ref{unitaryF}), leading
to the same eigenvalue of $L_{0}.$ All of these states are in infinite
dimensional unitary representations of SU$\left(  d,1\right)  .$ After
applying the Virasoro constraints the solutions get rearranged into
representations of the overall Poincar\'{e} symmetry\footnote{The separate
SU$\left(  d,1\right)  $ of each single oscillator is not expected to survive
in string theory because the Virarosoro constraints couple all the modes,
including the center of mass mode, to each other. Certainly there is at least
an overall Poincar\'{e} symmetry, and the states get rearranged into
representations of Poincar\'{e} with its little group (e.g. SO$\left(
d\right)  $ for massive states). Of course then the infinite dimensional
SU$\left(  d,1\right)  $ representations dissociate (they already are in the
SU$\left(  d\right)  \times$U$\left(  1\right)  $ basis in Eq.(\ref{towers}))
and rearranged properly according to Poincar\'{e} (or a larger hidden symmetry
if any such thing remains). \label{poincare}}. The good thing is that there
are no ghosts at all in this Fock space. However, it is not straightforward to
solve the Virasoro constraints for string states built on $|k,\tilde{0}%
\rangle$ because the creation-annihilation operators in the time direction
$\alpha_{n}^{0},\alpha_{-n}^{0}$ have their roles inverted while those in the
space directions $\vec{\alpha}_{n},\vec{\alpha}_{-n}$ remain the same.
Solutions seem likely to exist but none are known at this stage. If solutions
of the Virasoro constraints can be exhibited they would be of great interest
in string theory. This seems to be a challenging problem that we leave to
future work.

\begin{acknowledgments}
The current investigation was initiated after getting intrigued by a
conference lecture on the relativistic harmonic oscillator by Martin Land. I
would like to thank him for providing a copy of his transparencies
\cite{land}. I also thank L. Horwitz, K. Pilch and J. Schwarz, for their
constructive comments and discussions.
\end{acknowledgments}

\appendix

\section{SO$\left(  1,1\right)  $ oscillator in position space}

In this appendix we solve the differential equation $\left(  -\frac{1}%
{2}\partial^{\mu}\partial_{\mu}+\frac{1}{2}x^{\mu}x_{\mu}\right)
\psi_{\lambda}\left(  x\right)  =\lambda\psi_{\lambda}\left(  x\right)  $ in
the purely spacelike region\footnote{There are more general Lorentz covariant
solutions that have different forms in various spacelike and timelike regions
with continuity conditions across the lightcone $x^{\mu}x_{\mu}=0$ in Fig.1.
This will become evident in the discussion following Eq.(\ref{M}). For this
kind of solution the setting in section (\ref{unitaryF}) is more convenient.
In this section we will seek solutions with support only in the spacelike
regions, because those are the only ones described by the standard SO$(d,1)$
covariant Fock space approach discussed in section (\ref{vacsymm}), to which
we compare the solutions in this Appendix. \label{regions}} and show that we
arrive at the same conclusion as the oscillator approach using the Fock space
methods of section (\ref{symmvac}). For simplicity we will concentrate on
1-space and 1-time dimensions. Therefore the Lorentz symmetry is SO$\left(
1,1\right)  $ while the larger hidden symmetry is SU$\left(  1,1\right)  $.

We will discuss the spacelike region shown in Fig.1, knowing that the timelike
region is similar as indicated in section (\ref{symmvac}). Accordingly we
parametrize $x^{\mu}$ as follows to insure spacelike $x^{\mu}$%
\begin{equation}%
\begin{array}
[c]{c}%
x^{0}=\left\vert x\right\vert \sinh\theta,\;\;x^{1}=x\cosh\theta,\\
\text{both}~x,\theta\text{ range from }-\infty\text{ to }+\infty
\end{array}
\;\;\label{param}%
\end{equation}
This parametrization matches the parabolas in Fig.1 for fixed positive or
negative values of $x$, and as $x$ is varied the entire spacelike region is
covered. The differentials%
\begin{align}
dx^{0} &  =\varepsilon\left(  x\right)  \sinh\theta dx+\left\vert x\right\vert
\cosh\theta d\theta,\;\;\;dx^{1}=\cosh\theta dx+x\sinh\theta d\theta\\
dx &  =-\varepsilon\left(  x\right)  \sinh\theta dx^{0}+\cosh\theta
dx^{1},\;\;d\theta=\frac{\cosh\theta~dx^{0}}{\left\vert x\right\vert }%
-\frac{\sinh\theta~dx^{1}}{x}%
\end{align}
where $\varepsilon\left(  x\right)  \equiv sign\left(  x\right)  ,$ are useful
to compute the derivatives by using the chain rule $\frac{\partial}{\partial
x^{\mu}}=\frac{\partial\theta}{\partial x^{\mu}}\partial_{\theta}%
+\frac{\partial x}{\partial x^{\mu}}\partial_{x},$ to obtain
\begin{equation}
\frac{\partial}{\partial x^{0}}=\varepsilon\left(  x\right)  \left[
\frac{\cosh\theta}{x}\partial_{\theta}-\sinh\theta\partial_{x}\right]
,\;\;\frac{\partial}{\partial x^{1}}=-\frac{\sinh\theta}{x}\partial_{\theta
}+\cosh\theta\partial_{x}.
\end{equation}
The SO$\left(  1,1\right)  $ boost generator becomes (note extra sign due to
raising/lowering the timelike index $p^{0}=-i\partial/\partial x_{0}%
=+i\partial/\partial x^{0}$)%
\begin{equation}
L^{01}=x^{0}p^{1}-x^{1}p^{0}=-ix^{0}\frac{\partial}{\partial x^{1}}%
-ix^{1}\frac{\partial}{\partial x^{0}}=-i\varepsilon\left(  x\right)
\partial_{\theta}.
\end{equation}
%

%TCIMACRO{\FRAME{dtbpFU}{2.5399in}{1.6959in}{0pt}{\Qcb{Fig.1- Parabolas in the
%spacelike region of $\left(  x^{0},x^{1}\right)  $ \linebreak at some fixed
%$x=\pm a$ and any $\theta.$}}{}{oscillatorspacetime.eps}%
%width 2.5399in;  height 1.6959in;  depth 0pt;  original-width 2.4993in;
%original-height 1.6587in;  cropleft "0";  croptop "1";  cropright "1";
%cropbottom "0";  filename 'oscillatorSpacetime.eps';file-properties "XNPEU";}%
%}}%
%BeginExpansion
\begin{center}
\includegraphics[
height=1.6959in,
width=2.5399in
]%
{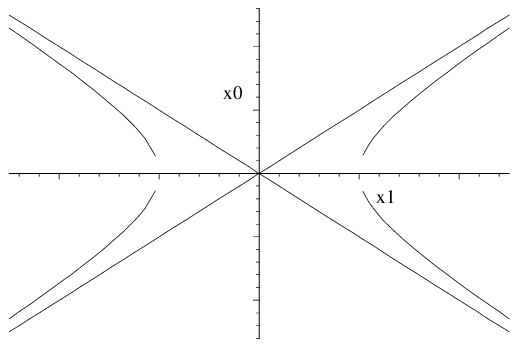}%
\\
Fig.1- Parabolas in the spacelike region of $\left(  x^{0},x^{1}\right)  $
\linebreak at some fixed $x=\pm a$ and any $\theta.$%
\end{center}
%EndExpansion
The operator $Q$ in $x^{\mu}$ space is then computed as%
\begin{equation}
Q=\frac{1}{2}\left(  p\cdot p+x\cdot x\right)  =\frac{1}{2}\left[
-\partial_{x}^{2}-\frac{1}{x}\partial_{x}+\frac{1}{x^{2}}\partial_{\theta}%
^{2}\right]  +\frac{1}{2}x^{2}.
\end{equation}
The solution of the eigenvalue equation $Q\psi_{\lambda m}=\lambda
\psi_{\lambda m}$ takes the separable form%
\begin{equation}
\psi_{\lambda m}\left(  x,\theta\right)  =x^{-1/2}F_{\lambda m}\left(
x\right)  e^{im\theta},\label{psi}%
\end{equation}
where the factor of $x^{-1/2}$ is inserted for convenience. The eigenvalue $m$
of the operator $\left(  -i\partial_{\theta}\right)  $ must be real if
$L^{01}=-i\varepsilon\left(  x\right)  \partial_{\theta}$ is to be hermitian.
This condition on $m$ \textit{imposes unitarity} hence only positive norms are
possible (see footnote \ref{hermit}). The range of $m$ is the entire
continuous real line $-\infty<m<\infty.$ Then $F_{\lambda m}\left(  x\right)
$ satisfies%
\begin{equation}
\left\{  -\partial_{x}^{2}-\frac{m^{2}+\frac{1}{4}}{x^{2}}+x^{2}%
-2\lambda\right\}  F_{\lambda m}\left(  x\right)  =0.\label{1D}%
\end{equation}
This is a one dimensional problem with an effective potential that has an
attractive (negative) component
\begin{equation}
V_{eff}\left(  x\right)  =-\frac{m^{2}+\frac{1}{4}}{2x^{2}}+\frac{1}{2}x^{2}.
\end{equation}
$V_{eff}\left(  x\right)  $ is plotted in Fig.2. For this shape of potential
we expect that there are normalizable bound states$.$ We also need to define a
normalization and \textit{include in the spectrum only the normalizable
solutions} of this equation.%

%TCIMACRO{\FRAME{dtbpFU}{2.5953in}{1.7236in}{0pt}{\Qcb{Fig.2 - Dashed line is
%for $m=0,$ solid line is for $m\neq0.$}}{}{oscillatoreffectivepotential.eps}%
%width 2.5953in;  height 1.7236in;  depth 0pt;  original-width 2.5538in;
%original-height 1.6864in;  cropleft "0";  croptop "1";  cropright "1";
%filename 'oscillatorEffectivePotential.eps';file-properties "XNPEU";}}}%
%BeginExpansion
\begin{center}
\includegraphics[
height=1.7236in,
width=2.5953in
]%
{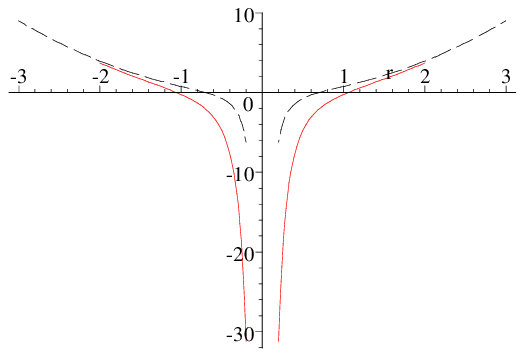}%
\\
Fig.2 - Dashed line is for $m=0,$ solid line is for $m\neq0.$%
\end{center}
%EndExpansion
We can choose the square integrable norm%
\begin{align}
\langle\psi_{\lambda m}|\psi_{\lambda^{\prime}m^{\prime}}\rangle &  =\int
d^{2}x\left(  \psi_{\lambda m}\left(  x\right)  \right)  ^{\ast}\psi
_{\lambda^{\prime}m^{\prime}}\left(  x\right) \label{square1}\\
&  =\int_{-\infty}^{\infty}dxF_{\lambda m}^{\ast}\left(  x\right)
F_{\lambda^{\prime}m^{\prime}}\left(  x\right)  \int_{-\infty}^{\infty}d\theta
e^{i\left(  m^{\prime}-m\right)  \theta}\\
&  =\delta\left(  m-m^{\prime}\right)  2\pi\int_{-\infty}^{\infty}dxF_{\lambda
m}^{\ast}\left(  x\right)  F_{k^{\prime}m^{\prime}}\left(  x\right) \\
&  =\delta\left(  m-m^{\prime}\right)  \delta_{kk^{\prime}} \label{squarenorm}%
\end{align}
In this case we must require a finite integral in $x$ space
\begin{equation}
2\pi\int_{-\infty}^{\infty}dxF_{km}^{\ast}\left(  x\right)  F_{k^{\prime}%
m}\left(  x\right)  =\delta_{kk^{\prime}}. \label{square2}%
\end{equation}

Next we solve for the allowed values of $k,m.$ The Schr\"{o}dinger equation in
Eq.(\ref{1D}) is related to the confluent hypergeometric equation and the
solutions are given by a linear superposition of the confluent hypergeometric
functions $M\left(  a,b,x^{2}\right)  ,U\left(  a,b,x^{2}\right)  .$ The
solution that is well behaved at $x^{2}\rightarrow\infty$ is given by%
\begin{equation}
\psi_{\lambda m}\left(  x\right)  =\alpha e^{-x^{2}/2}x^{im}U\left(  \left[
\frac{1}{2}-\frac{1}{2}\lambda+\frac{1}{2}im\right]  ,\left[  1+im\right]
,x^{2}\right)  ,\label{F}%
\end{equation}
where $\alpha$ is a normalization constant. This expression is even when $m$
is replaced by $-m$ due to the property $U\left(  a,b,z\right)  =z^{1-b}%
U\left(  1+a-b,2-b,z\right)  $. This is in agreement with the unitarity
condition of Eq.(\ref{T}), since the operator $T$ in that equation also
reverses the sign of the boost operator $L^{0i}\rightarrow-L^{0i},$ and hence
demands that only states that are even under $m\rightarrow-m$ can appear in
the unitary spectrum. Since we have already demanded unitarity of $L^{0i},$ it
has to be true that only states even under  $m\rightarrow-m$ should emerge
automatically in the spectrum of $Q$.

The behavior at $x\rightarrow\pm\infty$ is convergent $\psi_{\lambda m}\left(
x\right)  \sim\left\vert x\right\vert ^{\lambda-1}e^{-\frac{1}{2}x^{2}}\left(
1+O\left(  1/x^{2}\right)  \right)  $. The small $x\rightarrow0$ behavior is
given by (replace $m$ by $m\mp i\varepsilon$ with small real $\varepsilon$)%
\begin{equation}
\psi_{\lambda m}\left(  x\right)  \rightarrow\left\{
\begin{array}
[c]{l}%
\text{if }m\neq0:\frac{\Gamma\left(  1\pm im\right)  }{\Gamma\left(
\frac{1-\lambda}{2}\pm\frac{1}{2}im\right)  }\frac{\left\vert x\right\vert
^{\pm im}}{\pm im+\varepsilon}x^{\mp2\varepsilon},\;\varepsilon\rightarrow
0^{+},\\
\text{if }m=0:\frac{-1}{\Gamma\left(  \frac{1}{2}-\frac{1}{2}\lambda\right)
}\left(  \ln x^{2}+O\left(  1\right)  \right)  .
\end{array}
\right.  \label{M}%
\end{equation}
Therefore, the norm $\int dx\left\vert x\right\vert \left\vert \psi_{\lambda
m}\left(  x\right)  \right\vert ^{2}$ is integrable at $x=0,\pm\infty,$ hence
$\psi_{\lambda m}\left(  x\right)  $ is normalizable. This is in line with
expectations on the basis of the shape of the effective potential in Fig.2.

The probability density $\left\vert \psi_{\lambda m}\left(  x,\theta\right)
\right\vert ^{2}$ does not generally vanish at $x=0,$ which is everywhere at
the lightcone $x^{\mu}x_{\mu}=0$ in Fig.1. The physical meaning of this result
is that the oscillating particle in a spacelike region has generally a
non-vanishing probability at the lightcone. A similar computation in the
timelike region will also show that the lightcone is an allowed region of
spacetime. Therefore it would make sense to match the probability amplitude in
the spacetime region to the one in the timelike region at the lightcone. Then
we would get solutions in which the oscillating particle moves easily from the
spacetime to the timelike regions and vice versa. This kind of general
solution will be discussed in a more convenient setting in section
(\ref{unitaryF}).

There are however quantum states in which the leakage from the spacetime to
the timelike regions do not occur at all. This is seen by examining
Eq.(\ref{M}) and noting that for Lorentz singlets $\left(  m=0\right)  $ the
probability amplitude vanishes at the lightcone when $\frac{1}{2}-\frac{1}%
{2}\lambda$ is a negative integer or zero$.$ Hence only for the following
quantized values of $m,\lambda$ it is consistent to have a purely spacelike
relativistic harmonic oscillator%
\begin{equation}
m=0,\text{ and }\lambda=1+2k,\;\text{with integer }k=0,1,2,3,\cdots.
\end{equation}
For these values of $\lambda$ the solution $U$ reduces to a polynomial as follows%
\begin{equation}
\psi_{k}=\tilde{\alpha}_{k}e^{-x^{2}/2}U\left(  -k,1,x^{2}\right)  =\alpha
_{k}e^{-x^{2}/2}L_{k}^{0}\left(  x^{2}\right)  , \label{psik}%
\end{equation}
where $L_{k}^{0}\left(  x^{2}\right)  $ is the Laguerre polynomial with
argument $x^{2}$.
\begin{equation}
L_{k}^{0}\left(  x^{2}\right)  =\sum_{m=0}^{k}\frac{\left(  -1\right)  ^{m}%
k!}{\left(  m!\right)  ^{2}\left(  k-m\right)  !}x^{2m},\;k=0,1,2,3,\cdots
\end{equation}
So, the probability density $x|\psi|^2$ vanishes at the lightcone.

This result in $d=1$ is in full agreement with the oscillator
approach of section (\ref{symmvac}) for general $d.$ The oscillator
method, which was valid only for the spacelike region, also yielded
only Lorentz singlets Eq.(\ref{singlets}) as the only positive norm
states in a unitary representation of the Lorentz group SO$\left(
d,1\right)  $. Furthermore, the eigenvalues of
$Q\rightarrow\lambda=$ $1+2k$ agree when specialized to $d=1$.

What happened to the finite dimensional Lorentz representations with ghosts
that showed up in the Fock space approach in section (\ref{symmvac})? Those
had emerged in Fock space by applying oscillators $\bar{a}_{\mu}$ on the
vacuum $|0\rangle.$ What do we get if we follow the same approach in position
space? To investigate this we start with the oscillators in the Cartesian
basis%
\begin{align}
a_{0} &  =\frac{1}{\sqrt{2}}\left(  -x^{0}+\frac{\partial}{\partial x^{0}%
}\right)  ,\;\bar{a}_{0}=\frac{1}{\sqrt{2}}\left(  -x^{0}-\frac{\partial
}{\partial x^{0}}\right)  \;\\
a_{1} &  =\frac{1}{\sqrt{2}}\left(  x_{1}+\frac{\partial}{\partial x^{1}%
}\right)  ,\;\bar{a}_{1}=\frac{1}{\sqrt{2}}\left(  x_{1}-\frac{\partial
}{\partial x^{1}}\right)
\end{align}
and transform them to $\left(  x,\theta\right)  $ basis as%
\begin{align}
a_{0} &  =\frac{\varepsilon\left(  x\right)  }{\sqrt{2}}\left(  -\sinh
\theta\left(  x+\partial_{x}\right)  +\frac{\cosh\theta}{x}\partial_{\theta
}\right)  ,\;\;\\
\bar{a}_{0} &  =\frac{\varepsilon\left(  x\right)  }{\sqrt{2}}\left(
-\sinh\theta\left(  x-\partial_{x}\right)  -\frac{\cosh\theta}{x}%
\partial_{\theta}\right)  \\
a_{1} &  =\frac{1}{\sqrt{2}}\left(  \cosh\theta\left(  x+\partial_{x}\right)
-\frac{\sinh\theta}{x}\partial_{\theta}\right)  ,\;\;\\
\bar{a}_{1} &  =\frac{1}{\sqrt{2}}\left(  \cosh\theta\left(  x-\partial
_{x}\right)  +\frac{\sinh\theta}{x}\partial_{\theta}\right)
\end{align}
Clearly, $a_{0},a_{1}$ both annihilate the ground state $\psi_{vac}\left(
x,\theta\right)  =\langle x|0\rangle=e^{-x^{2}/2}$ since it is independent of
$\theta$ and satisfies $\left(  x+\partial_{x}\right)  e^{-x^{2}/2}=0$
\begin{equation}
a_{0}|0\rangle\rightarrow a_{0}e^{-x^{2}/2}=0,\;a_{1}|0\rangle\rightarrow
a_{1}e^{-x^{2}/2}=0.
\end{equation}
If we try to create states with the $\bar{a}_{1},\bar{a}_{0},$ we
automatically obtain solutions to the differential equation, but we see that
the $\theta$ dependence is not normalizable as follows%
\begin{align}
\bar{a}_{0}|0\rangle &  \Rightarrow\frac{\varepsilon\left(  x\right)  }%
{\sqrt{2}}\left(  -\sinh\theta\left(  x-\partial_{x}\right)  -\frac
{\cosh\theta}{x}\partial_{\theta}\right)  e^{-x^{2}/2}=-\sqrt{2}\left\vert
x\right\vert e^{-x^{2}/2}\sinh\theta\label{out1}\\
\bar{a}_{1}|0\rangle &  \Rightarrow\frac{1}{\sqrt{2}}\left(  \cosh
\theta\left(  x-\partial_{x}\right)  +\frac{\sinh\theta}{x}\partial_{\theta
}\right)  e^{-x^{2}/2}=\sqrt{2}xe^{-x^{2}/2}\cosh\theta\\
&  \text{These are solutions, but do not have the unitary form }e^{\pm
im\theta}.\label{out3}%
\end{align}
Indeed, the boost $L^{01}=-i\varepsilon\left(  x\right)  \partial_{\theta}$ is
hermitian only for the $e^{\pm im\theta}$ basis, it is not hermitian for the
$\left(  \sinh\theta,\cosh\theta\right)  $ or $e^{\pm\theta}$ basis.
Therefore, such excited states cannot be included in the spectrum if unitarity
is imposed from the beginning as was done in this section.

We emphasize that the oscillator states $\bar{a}_{0}|0\rangle,\bar{a}%
_{1}|0\rangle$ are excluded for two reasons. First, they are \textit{not in a
unitary representation of the Lorentz group} SO$\left(  1,1\right)  $ or of
the hidden symmetry group SU$\left(  1,1\right)  ,$ second they are
\textit{not normalizable} according to the square integrable norm defined
above because their norm diverges for the $\theta$ integral $\int_{-\infty
}^{\infty}d\theta\left(  \sinh\theta\right)  ^{2}=\infty,$ etc.. It is
important to emphasize that the square integrable norm above is different than
the Fock space norm. On that issue note that $\bar{a}_{0}|0\rangle,\bar{a}%
_{1}|0\rangle$ are normalizable if one uses the definition of norm in the
\textit{non-unitary} Fock space of section (\ref{vacsymm}), however this
admits negative as well as positive norms.

Following the oscillator approach in position space we obtain square
integrable normalizable states only for the singlets as follows. We
compute $\bar{a}\cdot\bar{a}$ and
note that it is independent of $\theta$%
\[
\bar{a}\cdot\bar{a}=-\bar{a}_{0}\bar{a}_{0}+\bar{a}_{1}\bar{a}_{1}=\frac{1}%
{2}\left(  x-\frac{1}{x}-\partial_{x}\right)  \left(  x-\partial_{x}\right)
.
\]
Therefore, $\left(  \bar{a}\cdot\bar{a}\right)  ^{k}$ creates $\theta
$-independent excited states, which are Lorentz singlets. For $k=1$ we can now
compute the oscillator state in Eq.(\ref{singlets}). This gives%
\begin{equation}
\left(  \bar{a}\cdot\bar{a}\right)  \langle x|0\rangle=\frac{1}{2}\left(
x-\frac{1}{x}-\partial_{x}\right)  \left(  x-\partial_{x}\right)  e^{-x^{2}%
/2}=2\left(  x^{2}-1\right)  e^{-x^{2}/2},
\end{equation}
which is in agreement with Eq.(\ref{psik}) for $k=1$%
\begin{equation}
\psi_{1}\left(  x\right)  =\alpha e^{-x^{2}/2}L_{1}^{0}\left(  x^{2}\right)
=\alpha e^{-x^{2}/2}\left(  1-x^{2}\right)  .
\end{equation}
More generally we can verify that the oscillator states $\left(  \bar{a}%
\cdot\bar{a}\right)  ^{k}\langle x|0\rangle$ reproduce the Laguerre
polynomials
\begin{align}
\psi_{k}\left(  x\right)   &  \sim\left(  \bar{a}\cdot\bar{a}\right)
^{k}\langle x|0\rangle\\
&  =\left[  \frac{1}{2}\left(  x-\frac{1}{x}-\partial_{x}\right)  \left(
x-\partial_{x}\right)  \right]  ^{k}e^{-x^{2}/2}\\
&  \sim\alpha_{k}e^{-x^{2}/2}L_{k}^{0}\left(  x^{2}\right)  .
\end{align}
These are certainly normalizable in $x$-space, and have positive norm, so they
are included in the positive norm spectrum. This is in complete agreement with
the results for general $d$ of section (\ref{symmvac}).

In the present approach the selection of the correct set of states emerged
automatically on the basis of normalizability and unitarity of the Lorentz
generator $L^{01}$ with the chosen norm of Eqs.(\ref{square1},\ref{square2}).
Of course, this amounts to the same criterion of section (\ref{symmvac}).

However, in the present approach we did not see so far why only the
vacuum state $\langle x|0\rangle$ must be kept. For this, we apply
the SU$\left(  1,1\right)  $ generators, such as $\bar{a}_{0}a_{1}$
or $\bar {a}_{1}a_{0}$ on the states $\psi_{\lambda m}\left(
x,\theta\right)  $ and note that this takes
us out of the unitary space $e^{im\theta}$ as explained in Eqs.(\ref{out1}%
-\ref{out3}). This means that the restriction to only the spacelike region,
\textit{plus unitarity}, breaks generally the SU$\left(  1,1\right)  $
covariance of the problem. This is like breaking symmetries via boundary
conditions. The covariance can be fully maintained only in the vacuum state.
Thus, if one is to seek solutions that are consistent with SU$\left(
1,1\right)  $ covariance, then only the vacuum state can satisfy this
criterion. Again this is in agreement with the Fock space approach of section
(\ref{vacsymm}).

\end{document}